\documentclass[preprint,aps,prd,nofootinbib,superscriptaddress]{revtex4}
\usepackage{hyperref}
\usepackage[pdftex]{graphicx,color}
\usepackage{slashed}
\usepackage{amsmath,amssymb}
\usepackage{hhline}
\usepackage{subfigure}
\usepackage{epstopdf}
\usepackage{epsfig}
\usepackage{graphicx}
\usepackage{caption}

\newcommand{\braket}[2]{\left\langle #1 | #2 \right\rangle}

\newcommand{\beq}{\begin{equation}}
\newcommand{\eeq}{\end{equation}}
\newcommand{\beqa}{\begin{eqnarray}}
\newcommand{\eeqa}{\end{eqnarray}}

\begin{document}
\begin{flushright}
WSU-HEP-1603\\
SI-HEP-2016-19
\end{flushright}
\title{\boldmath Lepton flavor violating quarkonium decays}

\author{Derek E. Hazard}
\affiliation{Department of Physics and Astronomy\\
        Wayne State University, Detroit, MI 48201, USA}

\author{Alexey A.\ Petrov}
\affiliation{Department of Physics and Astronomy\\
        Wayne State University, Detroit, MI 48201, USA}
\affiliation{Michigan Center for Theoretical Physics\\
        University of Michigan, Ann Arbor, MI 48196, USA}
\affiliation{Theoretische Physik 1, Naturwissenschaftlich-Technische 
Fakult\"at,\\
Universit\"at Siegen, D-57068 Siegen, Germany}


\begin{abstract}
\noindent
We argue that lepton flavor violating (LFV) decays $M \to \ell_1 \overline \ell_2$ of 
quarkonium states $M$ with different quantum numbers could be used to put constraints on 
the Wilson coefficients of effective operators
describing LFV interactions at low energy scales. We note that restricted kinematics of the 
two-body quarkonium decays allows us to select operators with particular quantum numbers, 
significantly reducing the reliance on the single operator dominance assumption that is
prevalent in constraining parameters of the effective LFV Lagrangian.
We shall also argue that studies of radiative lepton flavor violating 
$M \to \gamma \ell_1 \overline \ell_2$ decays could provide important complementary 
access to those effective operators.
\end{abstract}

\maketitle

\section{Introduction}\label{Intro}

Flavor-changing neutral current (FCNC) interactions serve as a powerful probe of physics beyond the standard 
model (BSM). Since no operators generate FCNCs in the standard model (SM) at tree level, new physics (NP) degrees of 
freedom can effectively compete with the SM particles running in the loop graphs, making their discovery possible.
This is, of course, only true provided the BSM models include flavor-violating interactions. 

The observation of charged lepton flavor violating (CLFV) transitions would provide especially clean probes of new 
physics. This is because in the standard model with massive neutrinos the CLFV transitions are suppressed by the 
powers of $m^2_\nu/m^2_W$, which renders the predictions for their transition rates vanishingly small, e.g. 
${\cal B}  (\mu \to e\gamma)_{\nu SM} \sim 10^{-54}$. A variety of well-established models on new physics
predict significantly larger rates for CLFV transitions \cite{Raidal:2008jk}. 

Any new physics scenario which involves lepton flavor violating interactions can be matched to an effective
Lagrangian, ${\cal L}_{\rm eff}$, whose Wilson coefficients would be determined by the ultraviolet (UV) physics
that becomes active at some scale $\Lambda$. 
Below the electroweak symmetry breaking scale, this Lagrangian must be invariant under unbroken 
$SU(3)_c\times U(1)_{\rm em}$ groups. The effective  
operators would reflect degrees of freedom relevant at the scale at which a given process takes place. If we 
assume that no new light particles (such as ``dark photons" or axions) exist in the low energy spectrum,
those operators would be written entirely in terms of the SM degrees of freedom such as leptons:
$\ell_i = \tau, \mu,$ and $e$; and quarks: $b, c, s, u,$ and $d$.
We shall not consider neutrinos in this paper. We also assume that top quarks have been integrated out.

The effective Lagrangian, ${\cal L}_{\rm eff}$, can then be divided into the dipole part, ${\cal L}_D$; 
a part that involves four-fermion interactions, ${\cal L}_{\ell q}$; and a gluonic part, ${\cal L}_{G}$.  
\begin{equation}\label{Leff}
{\cal L}_{\rm eff}= {\cal L}_D + {\cal L}_{\ell q} + {\cal L}_{G} + ... .
\end{equation}
Here the ellipses denote effective operators that are not relevant for the following analysis. 
The dipole part in Eq.~(\ref{Leff}) is usually written as \cite{Celis:2014asa}
\begin{eqnarray}\label{LD}
{\cal L}_{D} = -\frac{m_2}{\Lambda^2} \left[
\left( 
C_{DR}^{\ell_1\ell_2} \ \overline \ell_1 \sigma^{\mu\nu} P_L \ell_2 + 
C_{DL}^{\ell_1\ell_2} \ \overline \ell_1 \sigma^{\mu\nu} P_R \ell_2 
\right) F_{\mu\nu} + h.c. \right],
\end{eqnarray}
where $P_{\rm R,L}=(1\pm \gamma_5)/2$ is the right (left) chiral projection operator. The Wilson 
coefficients would, in general, be different for different leptons $\ell_i$. 

The four-fermion dimension-six lepton-quark Lagrangian takes the form:
\begin{eqnarray}\label{Llq}
{\cal L}_{\ell q} = -\frac{1}{\Lambda^2} \sum_q \Big[
\left( C_{VR}^{q\ell_1\ell_2} \ \overline\ell_1 \gamma^\mu P_R \ell_2 + 
C_{VL}^{q\ell_1\ell_2} \ \overline\ell_1 \gamma^\mu P_L \ell_2 \right) \ \overline q \gamma_\mu q &&
\nonumber \\
+ \
\left( C_{AR}^{q\ell_1\ell_2} \ \overline\ell_1 \gamma^\mu P_R \ell_2 + 
C_{AL}^{q\ell_1\ell_2} \ \overline\ell_1 \gamma^\mu P_L \ell_2 \right) \ \overline q \gamma_\mu \gamma_5 q &&
\nonumber \\
+ \
m_2 m_q G_F \left( C_{SR}^{q\ell_1\ell_2} \ \overline\ell_1 P_L \ell_2 + 
C_{SL}^{q\ell_1\ell_2} \ \overline\ell_1 P_R \ell_2 \right) \ \overline q q &&
\\
+ \
m_2 m_q G_F \left( C_{PR}^{q\ell_1\ell_2} \ \overline\ell_1 P_L \ell_2 + 
C_{PL}^{q\ell_1\ell_2} \ \overline\ell_1 P_R \ell_2 \right) \ \overline q \gamma_5 q 
\nonumber \\
+ \
m_2 m_q G_F \left( C_{TR}^{q\ell_1\ell_2} \ \overline\ell_1 \sigma^{\mu\nu} P_L \ell_2 + 
C_{TL}^{q\ell_1\ell_2} \ \overline\ell_1 \sigma^{\mu\nu} P_R \ell_2 \right) \ \overline q \sigma_{\mu\nu} q 
 &+& h.c. ~ \Big] .
\nonumber
\end{eqnarray}
We note that the tensor operators are often omitted when constraints on the Wilson coefficients in
Eq.~(\ref{Llq}) are derived (see, e.g. \cite{Celis:2014asa}). We would like to point out that those 
are no less motivated than others in Eq.~(\ref{Llq}). For example, they would be induced
from Fierz rearrangement of operators of the type 
${\cal Q} \sim \left(\overline q \ell_2\right)\left(\overline \ell_1 q \right)$ 
that often appear in leptoquark models.  Also, as we shall see later,  the experimental constraints on those 
coefficients follow from studying vector meson decays, where the best information on LFV transitions 
in quarkonia is available. 

The dimension seven gluonic operators can be either generated by some high scale physics or 
by integrating out heavy quark degrees of freedom \cite{Celis:2014asa,Petrov:2013vka},
\begin{eqnarray}\label{LG}
{\cal L}_{G} = -\frac{m_2 G_F}{\Lambda^2} \frac{\beta_L}{4\alpha_s} \Big[
\Big( C_{GR}^{\ell_1\ell_2} \ \overline\ell_1 P_L \ell_2 + 
C_{GL}^{\ell_1\ell_2} \ \overline\ell_1 P_R \ell_2 \Big)  G_{\mu\nu}^a G^{a \mu\nu} &&
\nonumber \\
+ ~ \Big( C_{\bar G R}^{\ell_1\ell_2} \ \overline\ell_1 P_L \ell_2 + 
C_{\bar G L}^{\ell_1\ell_2} \ \overline\ell_1 P_R \ell_2 \Big)  G_{\mu\nu}^a \widetilde G^{a \mu\nu}
 &+& h.c. \Big].
\end{eqnarray}
Here $\beta_L=-9 \alpha_s^2/(2\pi)$ is defined for the number of light active flavors, $L$, relevant to the scale 
of the process, which we take $\mu \approx 2$~GeV. All Wilson coefficients 
should also be calculated at the same scale. $G_F$ is the Fermi constant and 
$\widetilde G^{a \mu\nu} = (1/2) \epsilon^{\mu\nu\alpha\beta} G^a_{\alpha\beta}$ is a dual to the
gluon field strength tensor \cite{Celis:2014asa}.

The experimental constraints on the Wilson coefficients of effective operators in ${\cal L}_{\rm eff}$
could be obtained from a variety of LFV decays (see e.g. \cite{Raidal:2008jk} for a review). 
Deriving constraints on those Wilson coefficients usually involves an assumption that only one of the
effective operators dominates the result. This is not necessarily so in many particular UV 
completions of the LFV EFTs, so certain cancellations among contributions of various operators 
are possible. Nevertheless, single operator dominance is a useful theoretical assumption in placing
constraints on the parameters of ${\cal L}_{\rm eff}$.

In this paper we are going to argue that most of the Wilson coefficients of the effective Lagrangian in
Eq.~(\ref{Leff}) for different $\ell_i$ could be determined from experimental data on quarkonium decays. 
In particular, we consider two- and three-body decays of the quarkonia of differing quantum numbers 
with quarks of various flavors such as $\Upsilon(nS) \to  \ell_1 \overline \ell_2$, 
$\Upsilon(nS) \to \gamma  \ell_1 \overline \ell_2$, etc. We will note that restricted kinematics of the 
two-body transitions would allow us to select operators with particular quantum numbers significantly
reducing the reliance on the single operator dominance assumption. 
Finally, we shall argue that studies of radiative lepton flavor violating (RLFV) decays could provide 
important complementary access to study ${\cal L}_{\rm eff}$.

We shall provide calculations of the relevant decay rates and establish constraints, where
experimental data are available, on Wilson coefficients of effective operators of the Lagrangian 
${\cal L}_{\rm eff}$ of Eq.~(\ref{Leff}). 
In the following sections we assume CP-conservation, which implies that all Wilson coefficients will be treated 
as real numbers. We shall note that some transitions have not yet been experimentally studied, so no 
numerical constraints from those decays are available at the moment. Finally, in studying branching ratios we 
assume that for a meson, $M$, the branching fraction  
${\cal B}(M \to \ell_1\ell_2) = {\cal B}(M \to \overline \ell_1\ell_2) +{\cal B}(M \to \ell_1 \overline \ell_2)$, unless 
specified otherwise.

\section{Vector quarkonium decays $V \to \ell_1 \overline \ell_2$}\label{Spin1LL}

There is abundant experimental information on flavor off-diagonal leptonic decays of 
vector quarkonia, both from the ground and excited states \cite{PDG}. This information 
can be effectively converted to experimental bounds on Wilson coefficients of vector and 
tensor operators in Eq.~(\ref{Llq}), as well as on those of the dipole operators of Eq.~(\ref{LD}).
\begin{table*}
\caption{\label{tab:Vdecaylimits} Available experimental upper bounds on 
${\cal B}(V \to \ell_1 \ell_2)$ and ${\cal B}( \ell_2 \to \ell_1 \gamma)$ \cite{PDG}. 
Dashes signify that no experimental constraints are available and ``n/a" means 
that the transition is forbidden by available phase space. Charge averages of the final states are always assumed.}
\begin{ruledtabular}
\begin{tabular}{cccc}
$\ell_1 \ell_2$ &$\mu \tau$ & $e \tau$ & $e \mu$  \\ 
\hline
${\cal B}(\Upsilon (1S) \to \ell_1 \ell_2)$ & $ 6.0 \times 10^{-6}$ & $-$ & $-$  \\
${\cal B}(\Upsilon (2S) \to \ell_1 \ell_2)$ &  $3.3 \times 10^{-6}$ & $3.2 \times 10^{-6}$ & $-$ \\ 
${\cal B}(\Upsilon (3S) \to \ell_1 \ell_2)$ &  $3.1 \times 10^{-6}$ & $4.2 \times 10^{-6}$ & $-$ \\
${\cal B}(J/\psi \to \ell_1 \ell_2)$ &  $2.0 \times 10^{-6}$ & $8.3 \times 10^{-6}$ & $1.6 \times 10^{-7}$ \\
${\cal B}(\phi \to \ell_1 \ell_2)$ &  n/a & n/a & $4.1 \times 10^{-6}$ \\ 
${\cal B}(\ell_2 \to \ell_1 \gamma)$ & $4.4 \times 10^{-8}$ & $3.3 \times 10^{-8}$ & $5.7 \times 10^{-13}$ \\
\end{tabular}
\end{ruledtabular}
\end{table*}
The most general expression for the $V \to \ell_1 \overline \ell_2$ decay amplitude 
can be written as
\begin{widetext}
\begin{eqnarray}\label{Spin1Amp}
{\cal A}(V\to \ell_1 \overline \ell_2) = \overline{u}(p_1, s_1) \left[
A_V^{\ell_1\ell_2} \gamma_\mu + B_V^{\ell_1\ell_2} \gamma_\mu \gamma_5 
+ \frac{C_V^{\ell_1\ell_2}}{m_{V}} (p_2-p_1)_\mu 
\right. ~~~~~~~~~~~~~~
\nonumber \\
\qquad + \left.
\frac{iD_V^{\ell_1\ell_2}}{m_{V} }(p_2-p_1)_\mu \gamma_5 \
\right] v(p_2,s_2) \ \epsilon^\mu(p).
\end{eqnarray}
\end{widetext}
$A_V^{\ell_1\ell_2}$, $B_V^{\ell_1\ell_2}$, $C_V^{\ell_1\ell_2}$, and $D_V^{\ell_1\ell_2}$  are 
dimensionless constants which depend on the underlying Wilson coefficients of the effective 
Lagrangian of Eq.~(\ref{Leff}) as well as on hadronic effects associated with meson-to-vacuum 
matrix elements or decay constants.  

The amplitude of Eq.~(\ref{Spin1Amp}) leads to the branching fraction, which is convenient
to represent in terms of the ratio: 
\begin{eqnarray}\label{BRSpin1}
\frac{{\cal B}(V \to \ell_1 \overline \ell_2)}{{\cal B}(V \to e^+e^-)} &=& 
\left(\frac{m_V \left(1-y^2\right)}{4\pi\alpha  f_V Q_q}\right)^2 \Big[ \left(\left|A_V^{\ell_1\ell_2}\right|^2 +  \left|B_V^{\ell_1\ell_2}\right|^2\right)
+ \frac{1}{2} \left(1-2y^2\right)  \left(\left|C_V^{\ell_1\ell_2}\right|^2 +  \left|D_V^{\ell_1\ell_2}\right|^2\right) 
\nonumber \\
&+& y \ \text{Re}\left(A_V^{\ell_1\ell_2} C_V^{\ell_1\ell_2 *}+i B_V^{\ell_1\ell_2} D_V^{\ell_1\ell_2 *}\right) 
\Bigr] .
\end{eqnarray}
Here $\alpha$ is the fine structure constant, we neglected the mass of the lighter of the two leptons, 
and set $y=m_2/m_V$.  The form of the 
coefficients $A_V^{\ell_1\ell_2}$ to $D_V^{\ell_1\ell_2}$ depends on the initial state meson. For example, 
for $V=\Upsilon(nS)$ ($b\bar b$ states), 
$\psi(nS)$ ($c\bar c$ states), or $\phi$ ($s\bar s$ state), the coefficients are: 
\begin{eqnarray}\label{VCoef1}
A_V^{\ell_1\ell_2} &=& \frac{f_V m_V}{\Lambda^2} \left[ \ \
\sqrt{4\pi \alpha} Q_q y^2\left(C_{DL}^{\ell_1\ell_2}+C_{DR}^{\ell_1\ell_2}\right) + 
\kappa_V \left(C_{VL}^{q\ell_1\ell_2} + C_{VR}^{q\ell_1\ell_2}\right) \right.
\nonumber \\
&& \qquad\qquad + \left. 2 y^2 \kappa_V \frac{f^T_V}{f_V} G_F m_V m_q  
\left(C_{TL}^{q\ell_1\ell_2} + C_{TR}^{q\ell_1\ell_2}\right)
 \right],
\nonumber \\
B_V^{\ell_1\ell_2} &=& \frac{f_V m_V}{\Lambda^2} \left[
- \sqrt{4\pi \alpha} Q_q  y^2\left(C_{DL}^{\ell_1\ell_2}-C_{DR}^{\ell_1\ell_2}\right) - 
\kappa_V \left(C_{VL}^{q\ell_1\ell_2} - C_{VR}^{q\ell_1\ell_2}\right) \right.
\nonumber \\
&& \qquad\qquad - \left. 2 y^2 \kappa_V \frac{f^T_V}{f_V} G_F m_V m_q  
\left(C_{TL}^{q\ell_1\ell_2} -C_{TR}^{q\ell_1\ell_2}\right)
 \right],
\\
C_V^{\ell_1\ell_2} &=& \frac{ f_V m_V}{\Lambda^2} y \left[ \sqrt{4\pi \alpha} Q_q
\left(C_{DL}^{\ell_1\ell_2}+C_{DR}^{\ell_1\ell_2}\right) + 
2 \kappa_V \frac{f^T_V}{f_V}
G_F m_V m_q \left(C_{TL}^{q\ell_1\ell_2} + C_{TR}^{q\ell_1\ell_2}\right) \right],
\nonumber \\
D_V^{\ell_1\ell_2} &=& i \frac{ f_V m_V}{\Lambda^2} y \left[ -\sqrt{4\pi \alpha} Q_q 
\left(C_{DL}^{\ell_1\ell_2}-C_{DR}^{\ell_1\ell_2}\right) - 
2 \kappa_V \frac{f^T_V}{f_V}
G_F m_V m_q \left(C_{TL}^{q\ell_1\ell_2} - C_{TR}^{q\ell_1\ell_2}\right) \right].
\nonumber
\end{eqnarray}
Here $Q_q=(2/3, -1/3)$ is the charge of the quark $q$ and $\kappa_V = 1/2$ is a constant for pure 
$q\bar q$ states. It is a good approximation to drop terms proportional to $y^2$ in Eq.~(\ref{VCoef1})
for the heavy quarkonium states. Inspecting the ratio in Eq.~(\ref{BRSpin1}), one
immediately infers that the best constraints could be placed on the four-fermion coefficients, 
$C_{VL}^{q\ell_1\ell_2}$ and $C_{VR}^{q\ell_1\ell_2}$, as no final state lepton mass suppression
exists for those coefficients. Yet, constraints on the the dipole coefficients, 
$C_{DL}^{\ell_1\ell_2} (C_{DR}^{\ell_1\ell_2})$, are also possible in this case. This would provide NP
constraints that are complementary to the ones obtained from the lepton decay experiments, 
especially for $\ell = \tau$, obtained in the radiative $\tau \to \mu(e) \gamma$ decays.

The constraints on the Wilson coefficients of tensor operators, $C_{TL}^{q\ell_1\ell_2}(C_{TR}^{q\ell_1\ell_2})$,
in Eq.~(\ref{VCoef1}) also depend on the ratio of meson decay constants,
\begin{eqnarray}\label{DeConV}
\langle 0| \overline q \gamma^\mu q | V(p) \rangle &=& f_V m_V \epsilon^\mu (p)\,,
\nonumber \\
\langle 0| \overline q \sigma^{\mu\nu} q | V(p) \rangle &=& i f^T_V 
\left( \epsilon^\mu p^\nu-p^\mu \epsilon^\nu\right),
\end{eqnarray}
where $\epsilon^\mu(p)$ is the $V$-meson polarization vector, and $p$ is its momentum \cite{Becirevic:2013bsa}.

While the decay constants, $f_V$, are known, both experimentally from leptonic decays and 
theoretically from lattice or QCD sum rule calculations, for a variety of states $V$, the tensor 
(transverse) decay constant, $f^T_V$, has only recently been calculated for the charmonium $J/\psi$ 
state with the result $f^T_{J/\psi}(2\text{ GeV})=(410\pm10)$ MeV \cite{Becirevic:2013bsa}.
In the absence of the estimate for $f^T_V$, we follow the suggestion made in Ref.~\cite{Khodjamirian:2015dda}
and assume that $f^T_V=f_V$. This seems to be the case for the $J/\psi$ state \cite{Becirevic:2013bsa}
to better than 10 \%. We present numerical values of the decay constants in Table~\ref{tab:Vdecay_constants}.
Note that the ratio of Eq.~(\ref{BRSpin1}) is largely independent of the values of the decay constants. 

\begin{table}
\begin{center}
\footnotesize
\begin{tabular}{|c||c|c|c|c|c|c|c|}
\hline\hline
~State & $~\Upsilon(1S)~$ & $~\Upsilon(2S)~$ & $~\Upsilon(3S)~$ & $~J/\psi~$ & $~\psi(2S)$~  & $~\phi~$ & $~\rho \left(\omega \right)~$\\
\hline
\hline
~$f_V$, MeV~ &  $649\pm 31$   &  $481\pm 39$ &  $539\pm 84$ &  $418\pm 9$ &  $294\pm 5$ &  $241\pm 18$ & $209.4\pm 1.5$\\
\hline\hline
\end{tabular}
\normalsize
\end{center}
\caption{Vector meson decay constants used in the calculation of branching ratios ${\cal B}(V \to \ell_1 \overline \ell_2)$. 
The transverse decay constants are set $f^T_V=f_V$ except for $J/\psi$, which has 
$f^T_{J/\psi} = 410 \pm 10$ 
\cite{Colquhoun:2014ica,Abada:2015zea,Becirevic:2013bsa,MaiordeSousa:2012vv,Donald:2013pea,Chen:2015tpa}.
}\label{tab:Vdecay_constants} 
\end{table} 
Choosing other initial states would make it possible to constrain other combinations of the
Wilson coefficients in Eq.~(\ref{Leff}). This is important for the NP models where several LFV operators would 
contribute, especially in the case where no operator gives {\it a priori} dominant contribution. 
For example, choosing $V=\rho$ meson with $\rho \sim \left(u\bar u - d \bar d\right)/\sqrt{2}$ gives:
\begin{eqnarray}\label{VCoef2}
A_\rho^{e\mu} &=& \frac{f_\rho m_\rho}{\Lambda^2} y^2
\sqrt{2\pi \alpha} \left(Q_u - Q_d \right) \left(C_{DL}^{\ell_1\ell_2}+C_{DR}^{\ell_1\ell_2}\right),
\nonumber \\
B_\rho^{e\mu} &=& - \frac{f_\rho m_\rho}{\Lambda^2} y^2 \sqrt{2\pi \alpha} \left(Q_u - Q_d \right) \left(C_{DL}^{\ell_1\ell_2}-C_{DR}^{\ell_1\ell_2}\right),
\\
C_\rho^{e\mu} &=& \frac{ f_\rho m_\rho}{\Lambda^2} y \sqrt{2\pi \alpha} \left(Q_u - Q_d \right)
\left(C_{DL}^{\ell_1\ell_2}+C_{DR}^{\ell_1\ell_2}\right),
\nonumber \\
D_\rho^{e\mu} &=& -i \frac{ f_\rho m_\rho}{\Lambda^2} y \sqrt{2\pi \alpha} \left(Q_u - Q_d \right) 
\left(C_{DL}^{\ell_1\ell_2}-C_{DR}^{\ell_1\ell_2}\right).
\nonumber
\end{eqnarray}
Here we imposed isospin symmetry on the NP operators and their coefficients, which resulted 
in the cancellation of the four-fermion operator contribution. The restricted kinematics of the decay 
implies that only $\mu e$ operators can be constrained. The corresponding results for  
$V=\omega \sim \left(u\bar u + d \bar d\right)/\sqrt{2}$ decay can be obtained from Eq.~(\ref{VCoef1})
by substituting $Q_q \to \left(Q_u+Q_d\right)/\sqrt{2}$ and using $\kappa_{\omega} = 1/\sqrt{2} $. 
Again, the restricted kinematics of the decay implies 
that only $\mu e$ operators interacting with up and down quarks can be constrained. Since we 
imposed isospin symmetry, it is convenient to use $m_q = \left(m_u+m_d\right)/2$.

Contrasting Eq.~(\ref{BRSpin1}) with the experimental data from Ref.~\cite{PDG} we can constrain the 
Wilson coefficients of the Lagrangian Eq.~(\ref{Leff}). Assuming single operator dominance, the results 
can be found in Table \ref{tab:4fermion}. The Wilson coefficients of dipole operators can be found in 
Table \ref{tab:dipoles}.

\begin{table*}
\caption{\label{tab:4fermion}Constraints on the Wilson coefficients of four-fermion operators. Dashes signify that 
no experimental data are available to produce a constraint; ``n/a" means that the transition is forbidden by phase space.
Note that no experimental data is available for higher excitations of $\psi$.}
\begin{ruledtabular}
\begin{tabular}{ccccccc}
 & Leptons &\multicolumn{5}{c}{Initial state (quark)}\\
 Wilson coefficient ($GeV^{-2}$) & $\ell_1 \ell_2$ & $\Upsilon(1S) \ (b)$ & $\Upsilon(2S) \ (b)$ & $\Upsilon(3S) \ (b)$ 
 & $J/\psi \ (c)$ & $\phi \ (s)$  \\ \hline
$~$ & $\mu \tau$ & $ 5.6 \times 10^{-6}$ & $4.1 \times 10^{-6}$ & $3.5 \times 10^{-6}$ 
 & $5.5 \times 10^{-5}$ & n/a \\
$\left| C_{VL}^{q\ell_1\ell_2}/\Lambda^2 \right|$ & $e \tau$ & $-$ & $4.1 \times 10^{-6}$ & $4.1 \times 10^{-6}$ 
 & $1.1 \times 10^{-4}$ & n/a \\ 
$~$ & $e \mu$ & $-$ & $-$ & $-$  
& $1.0 \times 10^{-5}$ & $2 \times 10^{-3}$  \\
\hline
$~$ & $\mu \tau$  & $ 5.6 \times 10^{-6}$ & $4.1 \times 10^{-6}$ & $3.5 \times 10^{-6}$ 
 & $5.5 \times 10^{-5}$ & n/a \\
$\left| C_{VR}^{q\ell_1\ell_2}/{\Lambda^2} \right|$ & $e \tau$ & $-$ & $4.1 \times 10^{-6}$ & $4.1 \times 10^{-6}$ 
 & $1.1 \times 10^{-4}$ & n/a \\
$~$ & $e \mu$ & $-$ & $-$ & $-$ 
 & $1.0 \times 10^{-5}$ & $2 \times 10^{-3}$ \\
\hline
$~$  & $\mu \tau$  & $ 4.4 \times 10^{-2}$ & $3.2 \times 10^{-2}$ & $2.8 \times 10^{-2}$ 
 & $1.2$ & n/a \\
$\left| {C_{TL}^{q\ell_1\ell_2}}/{\Lambda^2} \right|$ & $e \tau$ & $-$ & $3.3 \times 10^{-2}$ & $3.2 \times 10^{-2}$ 
 & $2.4$ & n/a \\
$~$ & $e \mu$ & $-$ & $-$ & $-$ 
 & $4.8$ & $1 \times 10^{4}$ \\
\hline
$~$  & $\mu \tau$  & $ 4.4 \times 10^{-2}$ & $3.2 \times 10^{-2}$ & $2.8 \times 10^{-2}$ 
 & $1.2$ & n/a \\
$\left| {C_{TR}^{q\ell_1\ell_2}}/{\Lambda^2} \right|$ & $e \tau$ & $-$ & $3.3 \times 10^{-2}$ & $3.2 \times 10^{-2}$ 
 & $2.4$ & n/a \\
$~$ & $e \mu$ & $-$ & $-$ & $-$ 
 & $4.8$ & $1 \times 10^{4}$ \\
\end{tabular}
\end{ruledtabular}
\end{table*}

\begin{table*}
\caption{\label{tab:dipoles} Constraints on the dipole Wilson coefficients from the $1^{--}$ quarkonium
decays and radiative lepton transitions $\ell_2 \to \ell_1 \gamma$. Dashes signify that 
no experimental data are available to produce a constraint; ``n/a" means that the transition is 
forbidden by phase space.}
\begin{ruledtabular}
\begin{tabular}{cccccccc}
Dipole Wilson & Leptons &\multicolumn{5}{c}{Initial state} & \\
coefficient ($GeV^{-2}$) & $\ell_1 \ell_2$ & $\Upsilon(1S) \ (b)$ & $\Upsilon(2S) \ (b)$ & $\Upsilon(3S) \ (b)$ 
 & $J/\psi \ (c)$ & $\phi (s)$ & $\ell_2 \to \ell_1 \gamma$ \\ \hline
$~$ & $\mu \tau$ & $2.0 \times 10^{-4}$ & $1.6 \times 10^{-4}$ & $1.4 \times 10^{-4}$ 
 & $2.5 \times 10^{-4}$ & n/a & $2.6 \times 10^{-10}$ \\
$\left| C_{DL}^{\ell_1\ell_2}/\Lambda^2 \right|$ & $e \tau$ & $-$ & $1.6 \times 10^{-4}$ & $1.6 \times 10^{-4}$ 
 & $5.3 \times 10^{-4}$ & n/a & $2.7 \times 10^{-10}$ \\
$~$ & $e \mu$ & $-$ & $-$ & $-$  
& $1.1 \times 10^{-3}$ & $0.2$ & $3.1 \times 10^{-7}$ \\
\hline
$~$ & $\mu \tau$  & $2.0 \times 10^{-4}$ & $1.6 \times 10^{-4}$ & $1.4 \times 10^{-4}$ 
 & $2.5 \times 10^{-4}$ & n/a & $2.6 \times 10^{-10}$ \\
$\left| C_{DR}^{q\ell_1\ell_2}/{\Lambda^2} \right|$ & $e \tau$ & $-$ & $1.6 \times 10^{-4}$ & $1.6 \times 10^{-4}$ 
 & $5.3 \times 10^{-4}$ & n/a & $2.7 \times 10^{-10}$ \\
$~$ & $e \mu$ & $-$ & $-$ & $-$ 
 & $1.1 \times 10^{-3}$ & $0.2$ & $3.1 \times 10^{-7}$ \\
\end{tabular}
\end{ruledtabular}
\end{table*}

It is important to note that some of the bounds presented in Tables \ref{tab:4fermion} and \ref{tab:dipoles}
are rather weak and might not even look physically meaningful, especially the ones coming from 
$\phi$ decays. In fact, assuming Wilson coefficients  $C\sim 1$ seems to imply that new physics 
scale $\Lambda/\sqrt{C}$ only extends to several MeVs, clearly breaking the EFT paradigm that assumes
local operators up to the scales of several TeVs! A correct interpretation of those entries in 
Tables \ref{tab:4fermion} and \ref{tab:dipoles} is that existing data simply does not allow to place strong 
constraints on the combination Wilson coefficients. This is rather common in EFT analyses of new physics 
phenomena, see e.g. \cite{Petrov:2013nia}.

As one can see from Eq.~(\ref{VCoef1}), there is a practical limitation on the two-body 
vector meson decays.  Only a subset of the Wilson coefficients is selected by the 
quantum numbers of the initial state and can be probed. This fact can be turned into virtue if experimental 
information on LFV decays of quarkonium states with other quantum numbers is available.

\section{Pseudoscalar quarkonium decays $P \to \ell_1 \overline \ell_2$}\label{Spin0LL}

Constraints on other Wilson coefficients of the effective Lagrangian in Eq.~(\ref{Leff}) could be 
obtained by considering decays of pseudoscalar mesons with quantum numbers $0^{-+}$, 
which include states like $\eta_{b(c)}$, $\eta$, $\eta^{(\prime)}$, and their excitations. These decays 
would be sensitive to axial and pseudoscalar operators, providing information about 
$C_{PL}^{q\ell_1\ell_2}(C_{PR}^{q\ell_1\ell_2})$ and/or 
$C_{AL}^{q\ell_1\ell_2}(C_{AR}^{q\ell_1\ell_2})$ in Eq.~(\ref{Llq}) as well as to gluonic 
operators of Eq.~(\ref{LG}). The $\eta_{b(c)}$ states could be abundantly produced at 
the LHCb experiment directly in gluon-gluon fusion interactions \cite{Brambilla:2010cs}. 
In case of the $\eta_c$ and its excitations, another production mechanism would include 
non-leptonic $B$-decays, as the corresponding branching ratios for non-leptonic $B$ decays 
into $\eta_c$ and kaons are reasonably large, of order of per mille \cite{PDG}.
\begin{table*}
\caption{\label{tab:Pdecaylimits} Available experimental limits on ${\cal B}(P \to \ell_1 \ell_2)$ \cite{PDG}.
Note that no constraints for the heavy quark pseudoscalar states such as $\eta_{b(c)}$ are available. 
Only phase space allowed transitions are shown.}
\begin{ruledtabular}
\begin{tabular}{cc}
$\ell_1 \ell_2$ & $e \mu$  \\ 
\hline
${\cal B}(\eta \to \ell_1 \ell_2)$ & $6 \times 10^{-6}$ \\ 
${\cal B}(\eta^{\prime} \to \ell_1 \ell_2)$ & $4.7 \times 10^{-4}$ \\
${\cal B}(\pi^0 \to \ell_1 \ell_2)$ & $3.6 \times 10^{-10}$ \\
\end{tabular}
\end{ruledtabular}
\end{table*}

Similar to the decays of vector mesons considered in Sect.~\ref{Spin1LL}, one can write the most 
general expression for the $P \to \ell_1 \overline \ell_2$ decay amplitude as
\begin{eqnarray}\label{Spin0Amp}
{\cal A}(P\to \ell_1 \overline \ell_2) = \overline{u}(p_1, s_1) \left[
E_P^{\ell_1\ell_2}  + i F_P^{\ell_1\ell_2} \gamma_5 
\right] v(p_2,s_2) \,
\end{eqnarray}
with $E_P^{\ell_1\ell_2}$ and $F_P^{\ell_1\ell_2}$ being dimensionless constants which 
depend on the Wilson coefficients of operators in Eq.~(\ref{Leff}) and various decay constants.

The amplitude of Eq.~(\ref{Spin0Amp}) leads to the branching ratio for off-flavor diagonal leptonic 
decays of pseudoscalar mesons:
\begin{eqnarray}\label{BRSpin0}
{\cal B}(P \to \ell_1 \overline \ell_2) = \frac{m_P}{8\pi \Gamma_P} \left(1-y^2\right)^2
\left[\left|E_P^{\ell_1\ell_2}\right|^2 + \left|F_P^{\ell_1\ell_2}\right|^2\right].
\end{eqnarray}
Here $\Gamma_P$ is the total width of the pseudoscalar state.  We have once again 
neglected the mass of the lighter lepton and set $y = m_2/m_P$. 
Calculating $E_P^{\ell_1\ell_2}$ and $F_P^{\ell_1\ell_2}$  for $P=\eta_b$ ($b\bar b$ state) 
and $\eta_c$ ($c\bar c$ state), the coefficients are
\begin{eqnarray}\label{PCoef1}
E_P^{\ell_1\ell_2} &=& y \frac{m_P}{4 \Lambda^2} \left[
- i f_{P} \left[ 2 \left(C_{AL}^{q\ell_1\ell_2}+C_{AR}^{q\ell_1\ell_2}\right) - m_{P}^{2} G_{F} \left( C_{PL}^{q\ell_1\ell_2} + C_{PR}^{q\ell_1\ell_2} \right) \right] + 9 G_F a_{P} \left(C_{\widetilde G L}^{\ell_1\ell_2} + C_{\widetilde G R}^{\ell_1\ell_2} \right) \right],
\nonumber \\
F_P^{\ell_1\ell_2} &=& -y \frac{m_P}{4 \Lambda^2} \left[f_{P}\left[ 2 \left(C_{AL}^{q\ell_1\ell_2}-C_{AR}^{q\ell_1\ell_2}\right) - m_P^2 G_F \left(C_{PL}^{q\ell_1\ell_2}-C_{PR}^{q\ell_1\ell_2}\right) \right] + 9 i G_F a_{P} \left(C_{\widetilde G L}^{\ell_1\ell_2} - C_{\widetilde G R}^{\ell_1\ell_2} \right) \right].
\nonumber \\
\end{eqnarray}
The hadronic matrix elements in Eq.~(\ref{PCoef1}) are defined as \cite{Petrov:2013vka}
\begin{eqnarray}\label{DeConP}
&& \langle 0| \overline q \gamma^\mu \gamma_5 q | P(p) \rangle = -i f_P p^\mu\,,
\nonumber \\
&& \langle 0| \frac{\alpha_s}{4\pi} G^{a\mu\nu} \widetilde G^a_{\mu\nu}  | P(p) \rangle = a_P \,.
\end{eqnarray}
Here $p$ is the momentum of the meson. For heavy quarks $q=c,b$ one expects 
the matrix elements of gluonic operators in Eq.~(\ref{DeConP}) to be quite small\footnote{This can be visualized by 
noting that in the heavy quark limit $\eta_{b(c)}$ is a small state, of size $(m_{b(c)} v)^{-1}$, with small overlap with soft
gluons, whose Compton wavelength, of the order of $\Lambda_{\rm QCD}^{-1}$, is much larger than the distance 
between the quarks. Here $v$ is the velocity of heavy quarks.}, so we shall set $a_{\eta_{b(c)}}=0$ from now on.
The constraints on the Wilson coefficients of gluonic operators could be obtained either from studying 
lepton flavor violating $\eta^\prime$ decays (for $\mu e$ currents) or from the corresponding tau decays.
We use $a_{\eta} = -0.022 \pm 0.002$ GeV$^3$ and $a_{\eta^\prime} = -0.057 \pm 0.002$ GeV$^3$ \cite{Beneke:2002jn}.
The numerical values of the other pseudoscalar decay constants used in the 
calculations can be found in Table~\ref{tab:Pdecay_constants}.
\begin{table}
\begin{center}
\footnotesize
\begin{tabular}{|c||c|c|c|c|c|c|c|}
\hline\hline
~State & $~\eta_b~$ & $~\eta_c~$ & $\eta, u(d)$ & $~\eta, s$~  & $~\eta^\prime, u(d)~$  & $~\eta^\prime, s~$ & $~\pi~$\\
\hline
\hline
~$f_P^q$, MeV~ &  $667 \pm 6$   &  $387\pm 7$ &  $108\pm 3$ &  $-111\pm 6$ &  $89\pm 3$ &   $136\pm 6$&   $130.41 \pm 0.20$ \\
\hline\hline
\end{tabular}
\normalsize
\end{center}
\caption{Pseudoscalar meson decay constants used in the calculation of branching ratios ${\cal B}(P \to \ell_1 \overline \ell_2)$ \cite{McNeile:2012qf,Becirevic:2013bsa,Beneke:2002jn,PDG}.}
\label{tab:Pdecay_constants} 
\end{table} 
For the light quark states, such as $\eta$ and $\eta^\prime$ the corresponding expressions are a bit more involved:
\begin{eqnarray}\label{PCoef2}
E_P^{\ell_1\ell_2} &=& y \frac{m_P}{4 \Lambda^2} \left[
- i f_P^{u/d} \kappa_1^P \left[ 2 \left( C_{AL}^{{u/d}\ell_1\ell_2}+C_{AR}^{u/d\ell_1\ell_2} \right) - G_F m_P^2 \left( C_{PL}^{{u/d}\ell_1\ell_2}+C_{PR}^{u/d\ell_1\ell_2} \right) \right]  \right.
\nonumber \\
&-& \left. i f_P^s \kappa_2^P \left[ 2 \left( C_{AL}^{s \ell_1\ell_2}+C_{AR}^{s \ell_1\ell_2} \right) - G_F m_P^2 \left( C_{PL}^{s \ell_1\ell_2}+C_{PR}^{s \ell_1\ell_2} \right) \right]
+ 9 G_F a_{P} \left(C_{\widetilde G L}^{\ell_1\ell_2} + C_{\widetilde G R}^{\ell_1\ell_2} \right) \right],
\nonumber \\
F_P^{\ell_1\ell_2} &=& y \frac{m_P}{4 \Lambda^2} \left[
- f_P^{u/d} \kappa_1^P \left[ 2 \left( C_{AL}^{{u/d}\ell_1\ell_2}-C_{AR}^{u/d\ell_1\ell_2} \right) - G_F m_P^2 \left( C_{PL}^{{u/d}\ell_1\ell_2}-C_{PR}^{u/d\ell_1\ell_2} \right) \right]  \right.
\\
&-& \left. f_P^s \kappa_2^P \left[ 2 \left( C_{AL}^{s \ell_1\ell_2}-C_{AR}^{s \ell_1\ell_2} \right) - G_F m_P^2 \left( C_{PL}^{s \ell_1\ell_2}-C_{PR}^{s \ell_1\ell_2} \right) \right]
- 9 i G_F a_{P} \left(C_{\widetilde G L}^{\ell_1\ell_2} - C_{\widetilde G R}^{\ell_1\ell_2} \right) \right],
\nonumber 
\end{eqnarray}
where $\kappa_1^\eta= 1/\sqrt{3}$, $\kappa_2^\eta= -\sqrt{2/3}$, $\kappa_1^{\eta^\prime}= \sqrt{2/3}$, 
and $\kappa_2^{\eta^\prime}= 1/\sqrt{3}$.
It is important to note that, if observed, simultaneous fit to several light quark meson decays could independently constrain 
Wilson coefficients of effective operators in Eq.~(\ref{Leff}), as follows from Eq.~(\ref{PCoef2}).

\begin{table*}
\caption{\label{tab:ps_constr}Constraints on the Wilson coefficients from pseudoscalar meson decays. Dashes signify that 
no experimental data is available to produce a constraint; ``n/a" means that the transition is forbidden by phase space.}
\begin{ruledtabular}
\begin{tabular}{cccccccc}
 & Leptons &\multicolumn{6}{c}{Initial state}\\
 Wilson coefficient & $\ell_1 \ell_2$ & $\eta_b$ & $\eta_c$ & $\eta (u/d)$ & $\eta (s)$ & $\eta^\prime (u/d)$ & $\eta^\prime (s)$\\ \hline
$~$ & $\mu \tau$ & $-$ & $-$ & n/a & n/a & n/a & n/a\\
$\left| {C_{AL}^{q\ell_1\ell_2}}/{\Lambda^2} \right|$ & $e \tau$ & $-$ & $-$ & n/a & n/a & n/a & n/a \\
$~$ & $e \mu$ & $-$ & $-$ & $3 \times 10^{-3}$ & $2 \times 10^{-3}$ & $2.1 \times 10^{-1}$ & $1.9 \times 10^{-1}$\\
\hline
$~$ & $\mu \tau$ & $-$ & $-$ & n/a & n/a & n/a & n/a\\
$\left| {C_{AR}^{q\ell_1\ell_2}}/{\Lambda^2} \right|$ & $e \tau$ & $-$ & $-$ & n/a & n/a & n/a & n/a\\
$~$ & $e \mu$ & $-$ & $-$ & $3 \times 10^{-3}$ & $2 \times 10^{-3}$ & $2.1 \times 10^{-1}$ & $1.9 \times 10^{-1}$\\
\hline
$~$ & $\mu \tau$ & $-$ & $-$ & n/a & n/a & n/a & n/a\\
$\left| {C_{PL}^{q\ell_1\ell_2}}/{\Lambda^2} \right|$ & $e \tau$ & $-$ & $-$ & n/a & n/a & n/a & n/a\\
$~$ & $e \mu$ & $-$ & $-$ & $2 \times 10^{3}$ & $1 \times 10^{3}$ & $3.9 \times 10^{4}$ & $3.6 \times 10^{4}$\\
\hline
$~$ & $\mu \tau$ & $-$ & $-$ & n/a & n/a & n/a & n/a\\
$\left| {C_{PR}^{q\ell_1\ell_2}}/{\Lambda^2} \right|$ & $e \tau$ & $-$ & $-$ & n/a & n/a & n/a & n/a\\
$~$ & $e \mu$ & $-$ & $-$ & $2 \times 10^{3}$ & $1 \times 10^{3}$ & $3.9 \times 10^{4}$ & $3.6 \times 10^{4}$\\
\end{tabular}
\end{ruledtabular}
\end{table*}

\begin{table*}
\caption{\label{tab:gluon} Constraints on the pseudoscalar gluonic Wilson coefficients. Dashes signify that 
no experimental data is available to produce a constraint. No data for other lepton species is available.}
\begin{ruledtabular}
\begin{tabular}{cccccc}
Gluonic Wilson & Leptons &\multicolumn{4}{c}{Initial state} \\
coefficient ($GeV^{-2}$) & $\ell_1 \ell_2$ & $\eta_b$ & $\eta_c$ & $\eta$ & $\eta^{\prime}$ \\ 
\hline
$\left| C_{GL}^{\ell_1\ell_2}/\Lambda^2 \right|$  & $e \mu$ & $-$ & $-$ & $2 \times 10^2$  & $5.0 \times 10^{3}$ \\
$\left| C_{GR}^{\ell_1\ell_2}/{\Lambda^2} \right|$ & $e \mu$ & $-$ & $-$ & $2 \times 10^2$ & $5.0 \times 10^{3}$ \\
\end{tabular}
\end{ruledtabular}
\end{table*}

The resulting constraints on the Wilson coefficients could be found in Tables \ref{tab:ps_constr} and \ref{tab:gluon}. 
Note that no experimental constraints on the $b$ and $c$ currents are available, as the corresponding 
transitions $\eta_{b(c)} \to \ell_1\overline \ell_2$ have not yet been experimentally studied. 
Also, constraints on Wislon coefficients of the gluonic operators  in Table \ref{tab:gluon} are significantly 
weaker than those available from tau decays \cite{Petrov:2013vka}. Finally, just as in Sect. \ref{Spin1LL}, large 
entries in the Tables \ref{tab:ps_constr} and \ref{tab:gluon} do not imply a breakdown of the EFT description 
of LFV decays, but signify that existing data does not allow us to place strong constraints on the combination 
of relevant Wilson coefficients.

\section{Scalar quarkonium decays $S \to  \ell_1 \overline \ell_2$}\label{Spin0scalarLL}

Scalar quarkonium decays would ideally allow one to probe the Wilson coefficients of the scalar quark density 
operators in Eq.~(\ref{Llq}). The corresponding $p$-wave states $\chi_{q0}$, where $q=b,c$ could be effectively produced 
either directly in gluon-gluon fusion at the LHC, or in the radiative decays of $\Upsilon (2S)$, $\Upsilon (3S)$, or 
corresponding $\psi$ states. It is important to note that the corresponding branching ratios for, say, 
$\psi(2S) \to \gamma\chi_{c0}$ are rather large, of the order of 10\%. Finally, they could also be 
produced in $B$-decays at flavor factories.

Since Wilson coefficients of other operators could be better probed in the processes discussed in 
Sect.~\ref{Spin1LL}-\ref{Spin0LL}, in this section we shall concentrate on the contributions of operators that
could not be probed in the decays of vector or pseudoscalar quarkonium states.

The most general expression for the $S \to \ell_1 \overline \ell_2$ decay amplitude looks exactly like 
Eq.~(\ref{Spin0Amp}), with obvious modifications for the scalar decay:
\begin{eqnarray}\label{Spin0AmpS}
{\cal A}(S\to \ell_1 \overline \ell_2) = \overline{u}(p_1, s_1) \left[
E_S^{\ell_1\ell_2}  + i F_S^{\ell_1\ell_2} \gamma_5 
\right] v(p_2,s_2) .
\end{eqnarray}
$E_S^{\ell_1\ell_2}$ and $F_S^{\ell_1\ell_2}$ are dimensionless constants. The branching ratio, which follows from 
Eq.~(\ref{Spin0AmpS}), is
\begin{eqnarray}\label{BRSpin0S}
{\cal B}(S \to \ell_1 \overline \ell_2) = \frac{m_S}{8\pi \Gamma_S} \left(1-y^2\right)^2
\left[\left|E_S^{\ell_1\ell_2}\right|^2 + \left|F_S^{\ell_1\ell_2}\right|^2\right].
\end{eqnarray}
Here $\Gamma_S$ is the total width of the scalar state and $y = m_2/m_S$. 
The coefficients $E_S^{\ell_1\ell_2}$ and $F_S^{\ell_1\ell_2}$  are
\begin{eqnarray}\label{SCoef1}
E_S^{\ell_1\ell_2} &=& y \frac{m_S G_F}{4 \Lambda^2} 
\left[2 i f_{S} m_S m_q \left(C_{SL}^{q l_1 l_2} + C_{SR}^{q l_1 l_2}\right) +
9 a_S \left(C_{GL}^{q l_1 l_2} + C_{GR}^{q l_1 l_2}\right) \right],
\nonumber \\
F_S^{\ell_1\ell_2} &=& y \frac{m_S G_F}{4 \Lambda^2}
 \left[2 f_{S} m_S m_q \left(C_{SL}^{q l_1 l_2} - C_{SR}^{q l_1 l_2}\right) -
9 i a_S \left(C_{GL}^{q l_1 l_2} - C_{GR}^{q l_1 l_2}\right)
\right].
\end{eqnarray}
The hadronic matrix elements in Eq.~(\ref{SCoef1}) are defined as
\begin{eqnarray}\label{DeConS}
&&  \langle 0| \overline q q | S(p) \rangle = -i m_S f_S \ ,
\nonumber \\
&& \langle 0| \frac{\alpha_s}{4\pi} G^{a\mu\nu} G^a_{\mu\nu}  | S(p) \rangle = a_S \,.
\end{eqnarray}
Note that we introduced an extra minus sign and a factor of $m_S$ compared to \cite{Godfrey:2015vda} 
for the scalar quark density to have uniform units for all matrix elements of quark currents. 
For the same reasons as in the pseudoscalar case, one expects that the gluonic 
matrix elements in Eq.~(\ref{SCoef1}) for the heavy quark states $\chi_{c0}$ or $\chi_{b0}$ are 
small, so we set $a_S=0$ from now on. This means that the Wilson coefficients of the gluonic
operators are better probed in LFV tau decays, where the low energy theorems \cite{Petrov:2013vka}
or experimental data \cite{Celis:2014asa} could be used to constrain relevant gluonic matrix elements. 

\begin{table}
\begin{center}
\footnotesize
\begin{tabular}{|c||c|c|c|c|c|c|c|c|}
\hline\hline
~State & $~\chi_{c0} (1P)~$ & $~\chi_{b0} (1P)~$ & $~\chi_{b0} (2P)~$ \\
\hline
\hline
~$m_S$, MeV~ &  $~3414.75\pm 0.31~$   &  $~9859.44\pm 0.52~$ &  $10232.5\pm 0.6~$ \\
~$\Gamma_S$, MeV~ &  $10.5\pm 0.6$   &  $(1.35)$ &  $~(0.247 \pm 0.097)~~$ \\
~$f_S$, MeV~ &  $887$   &  $423$ &  $421$ \\
\hline\hline
\end{tabular}
\normalsize
\end{center}
\caption{Decay constants of Eq.~(\ref{DeConP}) for the scalar quarkonium decays, derived from the
quark model calculations of \cite{Godfrey:2015vda}. Masses and measured widths are from \cite{PDG}, and 
unmeasured widths (in brackets) are calculated as in \cite{Godfrey:2015vda,Godfrey:2015dia}.
}\label{tab:BRradiative} 
\end{table} 

Finally, we note that no constraints on the Wilson coefficients of the scalar currents in ${\cal L}_{\rm eff}$ are 
available, as the corresponding transitions $\chi_{b(c)0} \to \ell_1 \overline \ell_2$ have not yet been 
experimentally studied. 

\section{Three body vector quarkonium decays $V \to  \gamma \ell_1 \overline \ell_2 $}\label{Spin1LLgam}

Addition of a photon to the final state certainly reduces the number of the events available for 
studies of LFV decays, especially since no compensating mechanisms seem to be present 
({\it c.f.} \cite{Aditya:2012ay}). However, it is also makes it possible for operators in ${\cal L}_{\rm eff}$,
other than considered in two-body decays, to contribute, which makes the analysis of RLFV decays a 
worthwhile exercise, especially for the decays of the vector quarkonium states.

\subsection{Resonant transitions}

The resonant two-body radiative transitions of vector states $V \to \gamma (M \to \ell_1 \overline \ell_2)$ could 
be used to study two-body decays considered above, provided the corresponding branching ratios for the
radiative decays are large enough. Since vector states are abundantly produced in $e^+e^-$ annihilation,
these decays could provide a powerful tool to study LFV transitions at flavor factories. 

If the soft photon can be effectively tagged at B-factories, the combined branching ratio factorizes and 
can be written as
\begin{equation}
{\cal B}(V \to \gamma  \ell_1 \overline \ell_2) = 
{\cal B}(V \to \gamma M) {\cal B}(M \to \ell_1 \overline \ell_2),
\end{equation} 
where the scalar decays ($M=\chi_{q0}$) ${\cal B}(\chi_{q0} \to  \ell_1 \overline \ell_2)$ have been studied
 in Sect.~\ref{Spin0scalarLL}, while the corresponding pseudoscalar transitions ($M=\eta_q$) 
 ${\cal B}(\eta_q \to  \ell_1 \overline \ell_2)$ have been studied in Sect.~\ref{Spin0LL}.

The resonant RLFV decays are quite useful for studies of scalar heavy meson decays, as the 
corresponding branching ratios are large, of order of a few percent \cite{PDG}. In charm,
\begin{eqnarray}\label{BranchRadc}
&&  {\cal B}(\psi(2S) \to \gamma \chi_{c0} (1P)) = 9.99 \pm 0.27\% \ ,
\nonumber \\
&& {\cal B}(\psi(3770) \to \gamma \chi_{c0}(1P)) = 0.73 \pm 0.09\% \ .
\nonumber 
\end{eqnarray}
The corresponding radiative transitions in beauty sector are also rather large,
\begin{eqnarray}\label{BranchRadb}
&& {\cal B}(\Upsilon(2S) \to \gamma \chi_{b0}(1P)) = 3.8 \pm 0.4\% \ ,
\nonumber \\
&& {\cal B}(\Upsilon(3S) \to \gamma \chi_{b0}(1P)) = 0.27 \pm 0.04\% \ ,
\\
&&  {\cal B}(\Upsilon(3S) \to \gamma \chi_{b0} (2P)) = 5.9 \pm 0.6\% \ .
\nonumber 
\end{eqnarray}
A rough estimate \cite{Godfrey:2015vda} shows that with the integrated luminosity of 
${\cal L}=250$ fb$^{-1}$ the number of produced $\chi_b$ states could reach tens of millions. 
Thus, studies of LFV transitions of $\chi_b$ states could result in a solid bound on the Wilson 
coefficients of the scalar operators in ${\cal L}_{\rm eff}$.

Similar radiative transitions to the pseudoscalar states are generally smaller. However, since the 
pseudoscalar $0^{-+}$ states are lighter than the $1^{--}$ ones, the radiative transition rates could still 
reach a percent level in charm:
\begin{eqnarray}\label{BranchRadPSc}
&& {\cal B}(J/\psi \to \gamma \eta_c) = 1.7 \pm 0.4\% \ ,
\nonumber \\
&& {\cal B}(\psi(2S) \to \gamma\eta_c) = 0.34 \pm 0.05\% \ .
\nonumber 
\end{eqnarray}
The corresponding branching ratios in $b$ sector are in a sub permille level and cannot be effectively used to 
study LFV decays of the $\eta_b$ states.

\subsection{Non-resonant transitions}

Non-resonant three-body radiative decays of vector states $V \to \gamma \ell_1 \overline \ell_2$ could be used to 
constrain the scalar operators, which are not accessible in the two-body decays of vector or pseudoscalar states.  
Since the final state now includes the photon, it is no longer possible to express all of the hadronic effects
in terms of the decay constants. The constraints would then depend on a set of $V \to \gamma$ form factors that
are not well known. We shall discuss those in a future publication \cite{HazardPetrovFuture}.

Here we would provide information about $C_{SL}^{q \ell_1 \ell_2} (C_{SR}^{q \ell_1 \ell_2})$, but at the expense of 
introducing model dependence. We shall calculate the transition $V \to \gamma \ell_1 \overline \ell_2$ choosing a
particular model to describe the effective quark-antiquark distribution function \cite{Aditya:2012ay}. 

In principle, besides the Wilson coefficients of the scalar operators, non-resonant radiative LFV decays could be 
used to obtain information about vector, axial, pseudoscalar, and tensor operators and thus 
$C_{VL}^{q \ell_1 \ell_2} (C_{VR}^{q \ell_1 \ell_2})$, $C_{AL}^{q \ell_1 \ell_2} (C_{AR}^{q \ell_1 \ell_2})$, 
$C_{PL}^{q \ell_1 \ell_2} (C_{PR}^{q \ell_1 \ell_2})$, and  $C_{TL}^{q \ell_1 \ell_2} (C_{TR}^{q \ell_1 \ell_2})$.  
However, because these operators can be constrained using much simpler two-body decays of vector and 
pseudoscalar states (see Sec.~\ref{Spin1LL}-\ref{Spin0LL}) without significant model dependence, and with better
statistics, we shall focus here mainly on the scalar operators, leaving the other constraints to the 
future work \cite{HazardPetrovFuture}.
\begin{figure}
\subfigure[]{\includegraphics[scale=0.45]{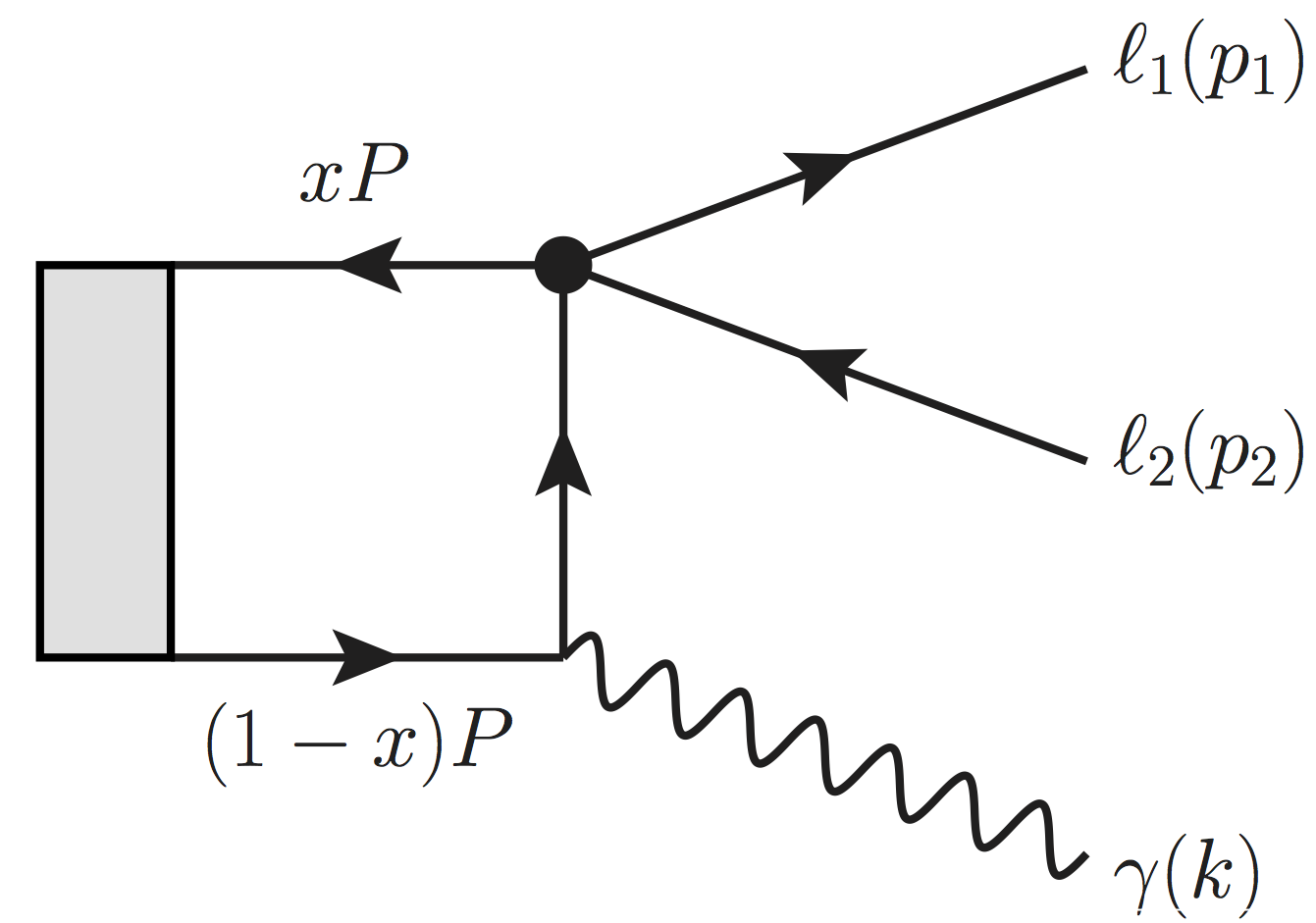} \label{diagram a}}
\subfigure[]{\includegraphics[scale=0.45]{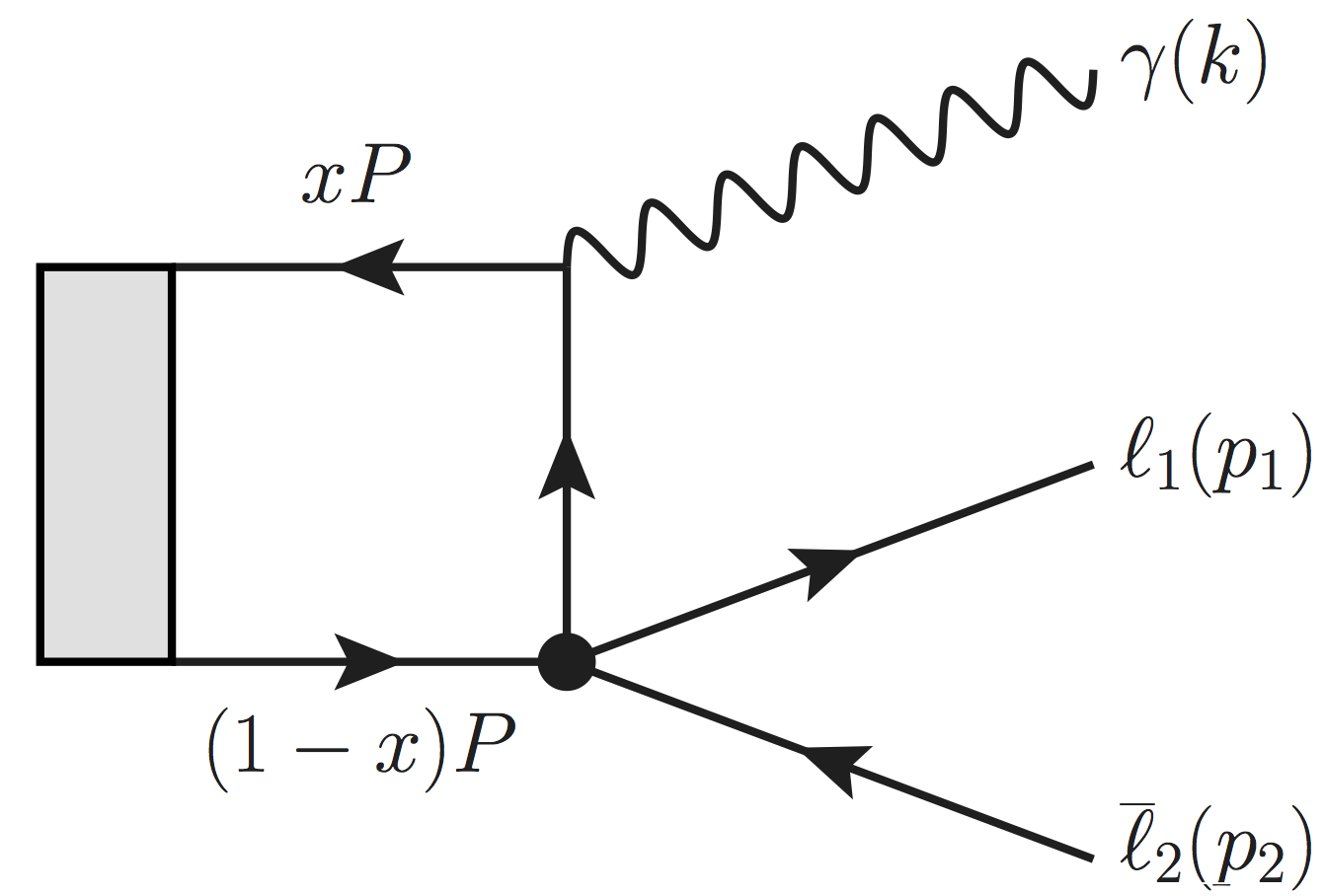} \label{diagram b}} \\
\subfigure[]{\includegraphics[scale=0.45]{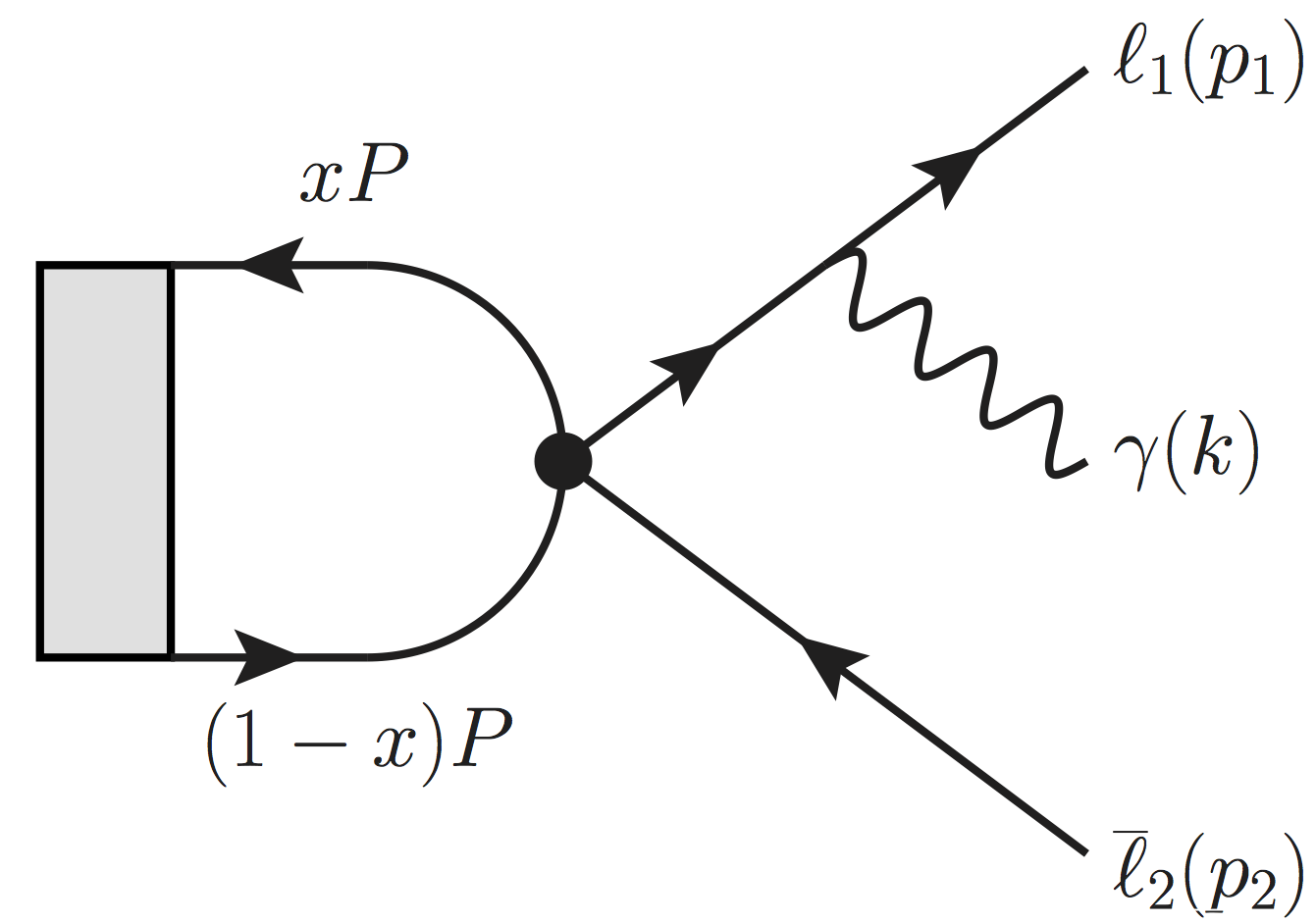} \label{diagram c}}
\subfigure[]{\includegraphics[scale=0.45]{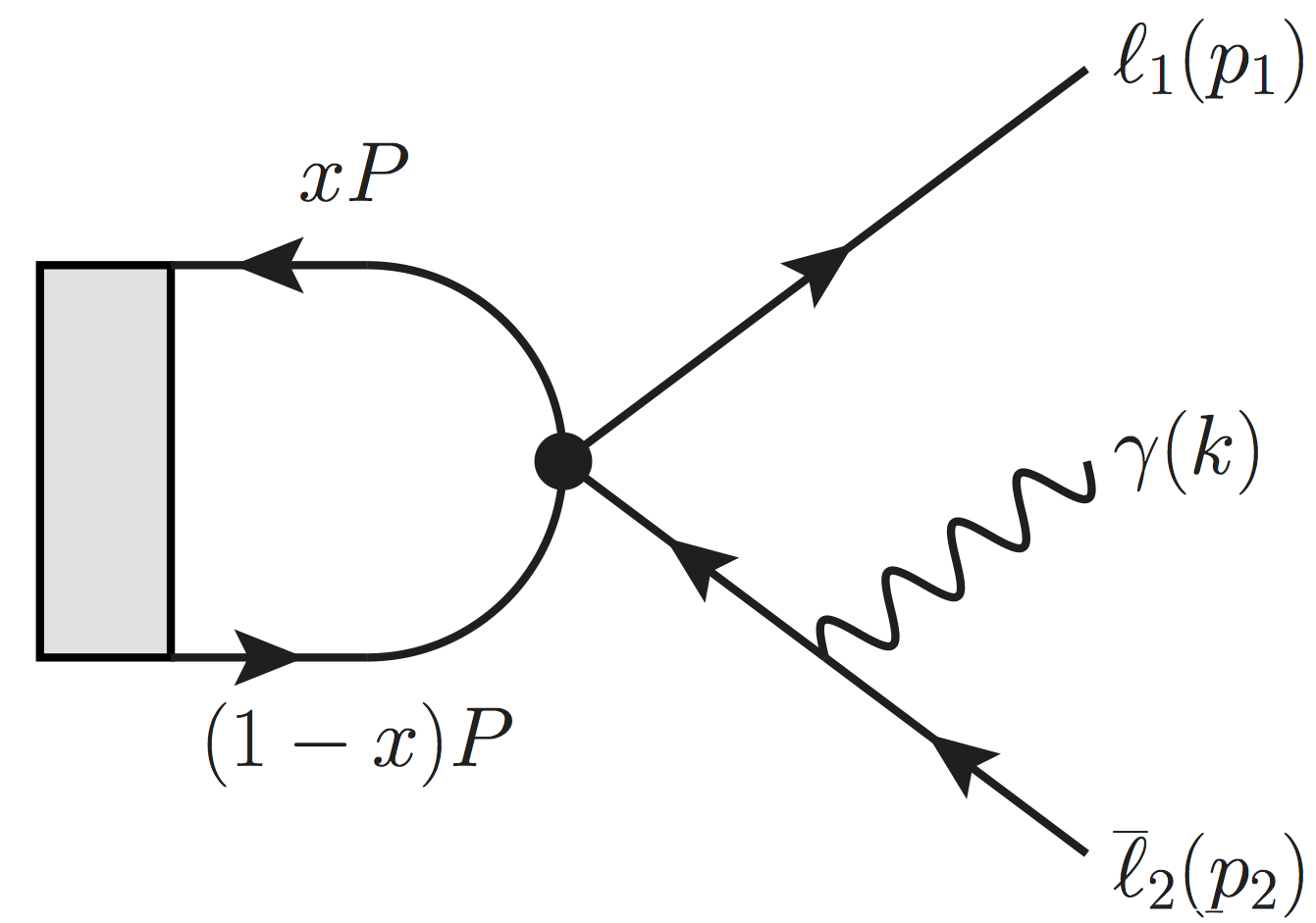} \label{diagram d}} \\
\subfigure[]{\includegraphics[scale=0.45]{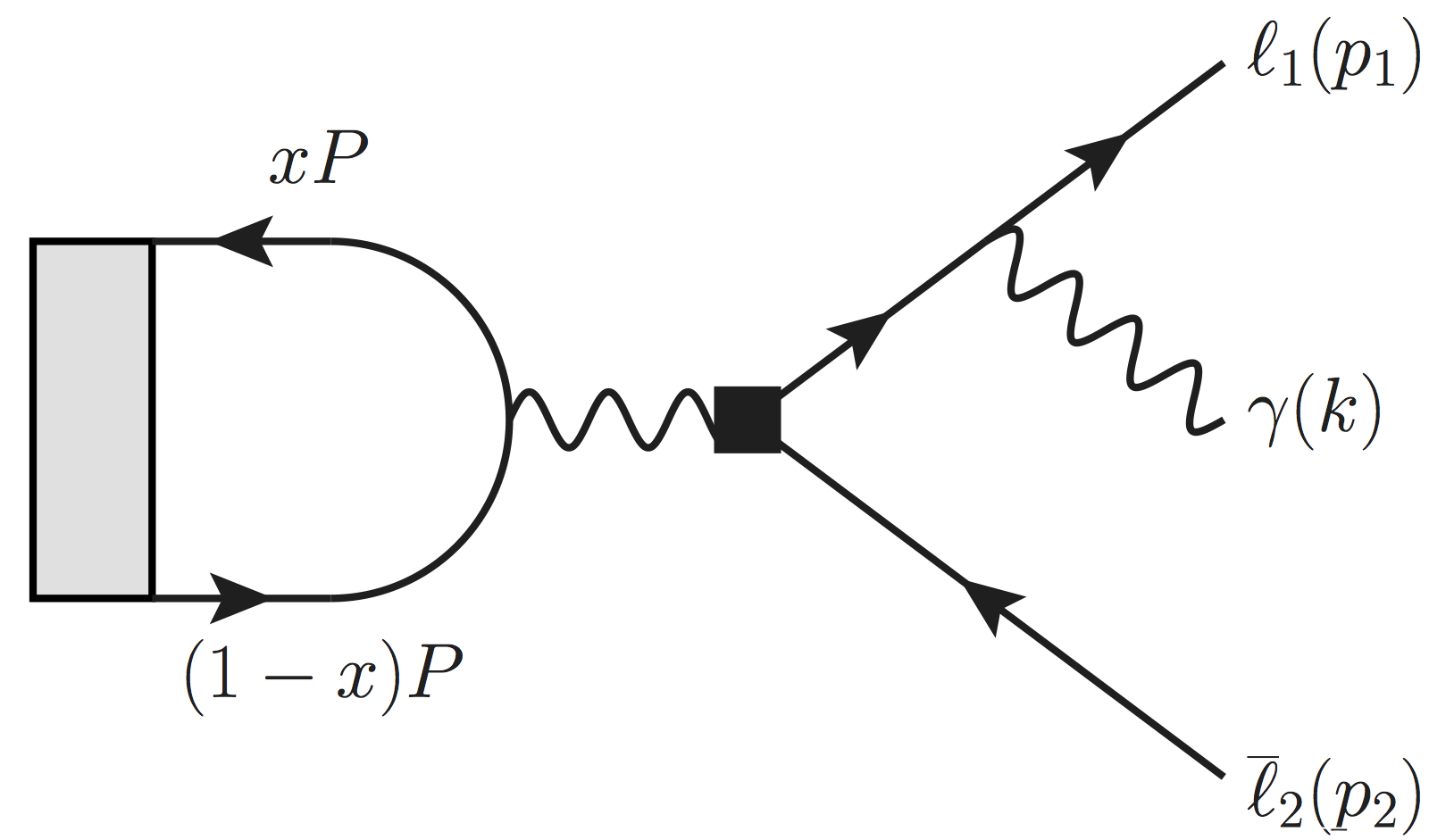} \label{diagram e}}
\subfigure[]{\includegraphics[scale=0.45]{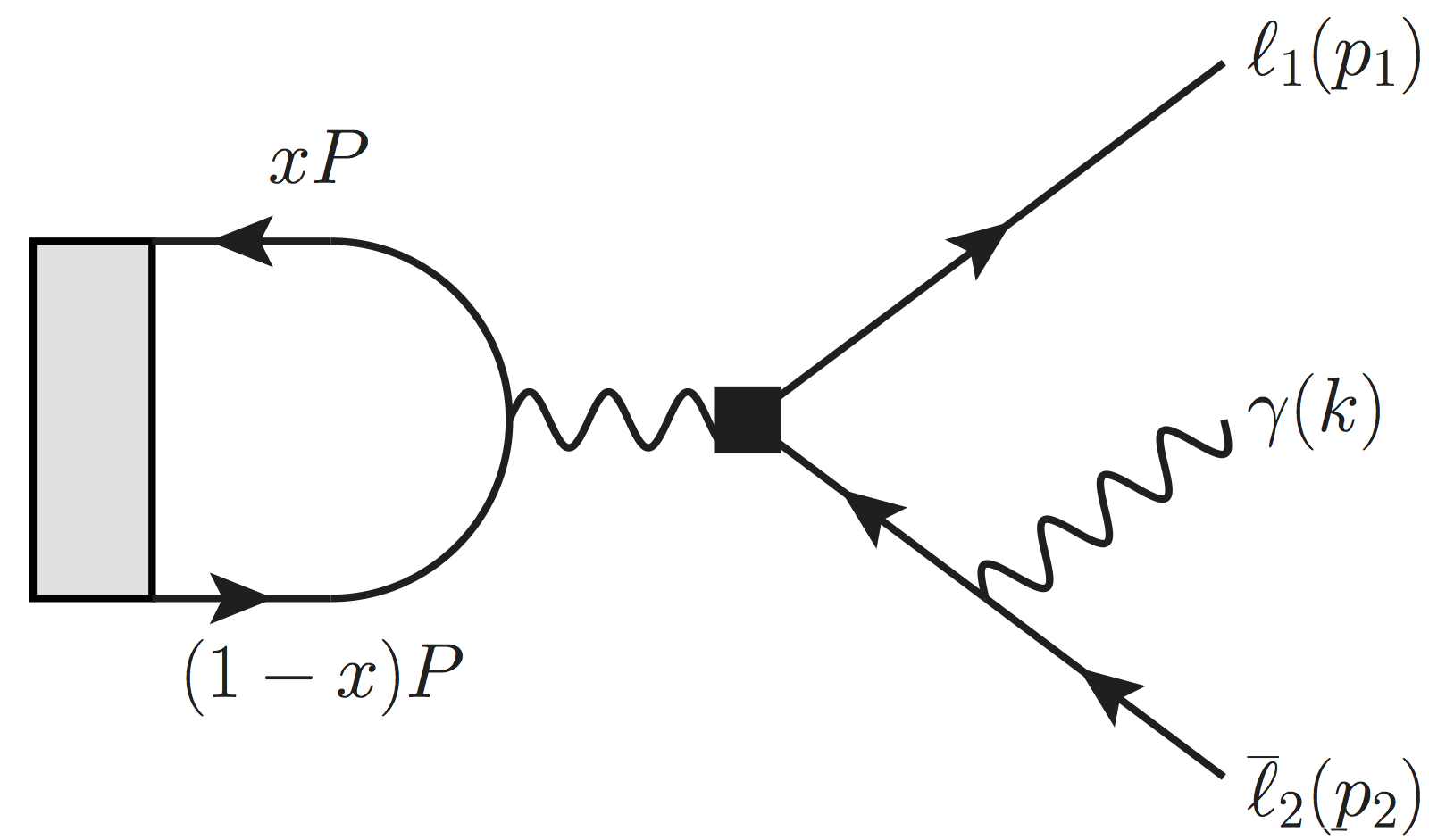} \label{diagram f}} \\
\subfigure[]{\includegraphics[scale=0.45]{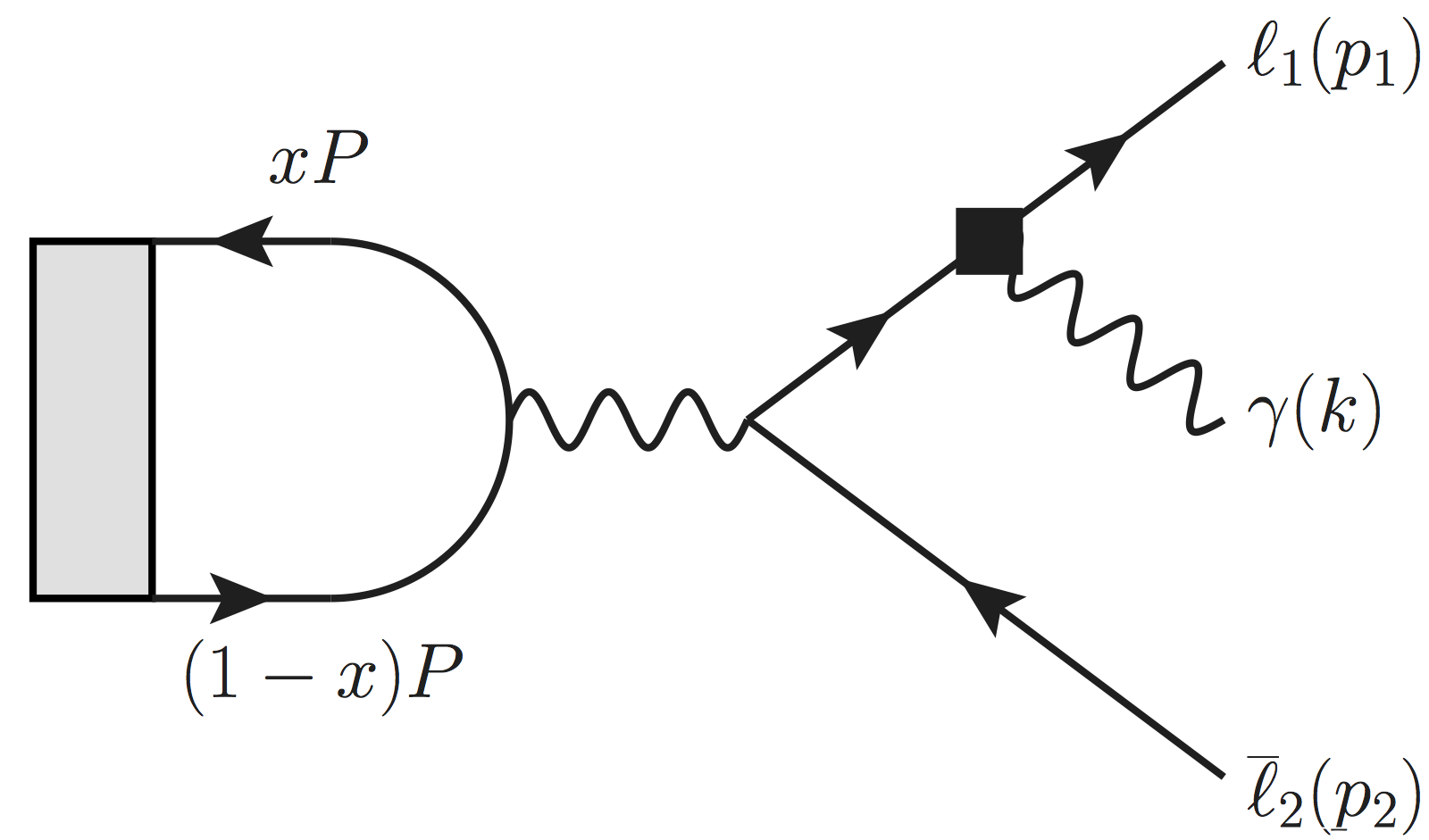} \label{diagram g}}
\subfigure[]{\includegraphics[scale=0.45]{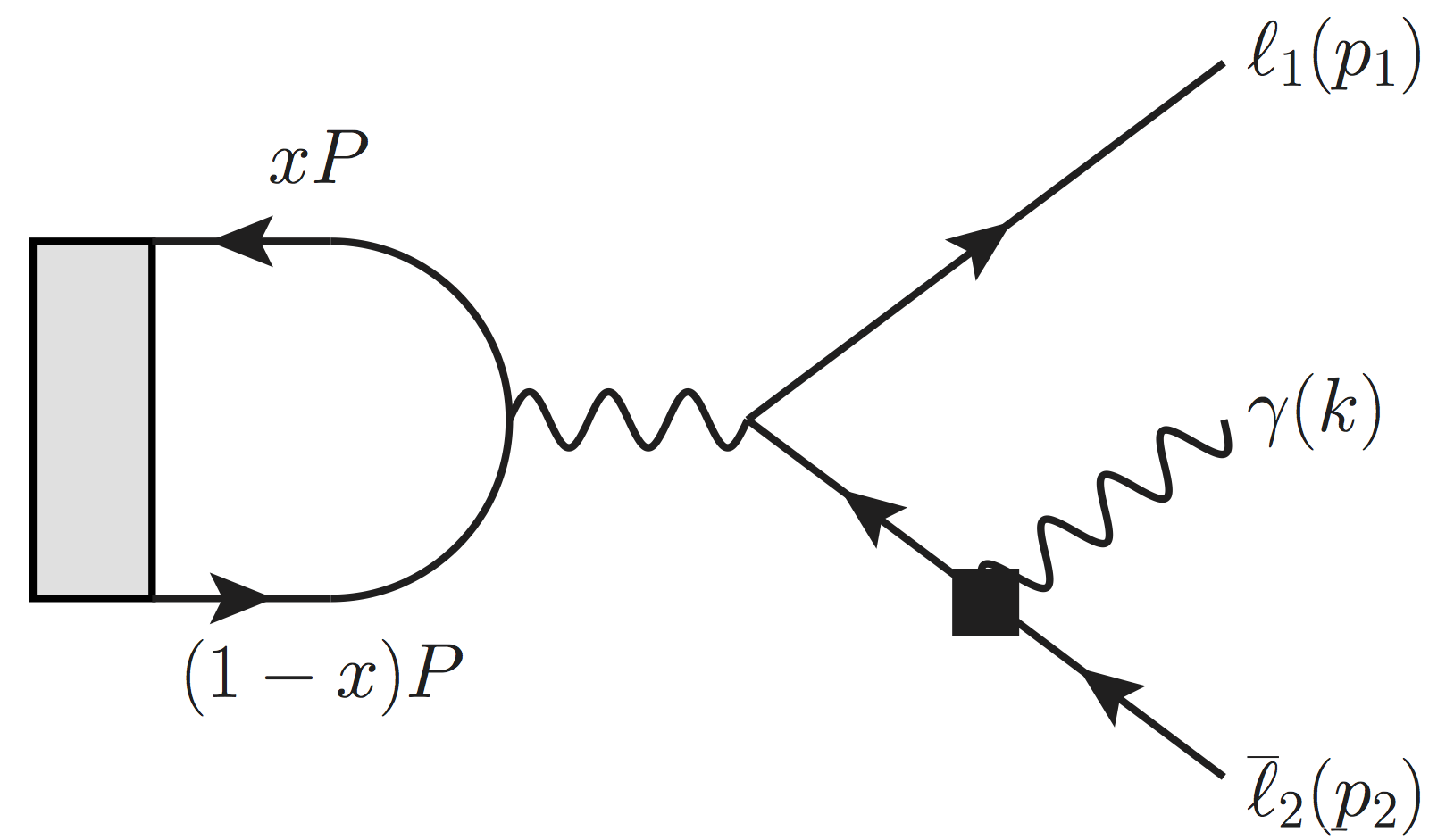} \label{diagram h}}
\caption{ Feynman diagrams for ${\cal A} (V \to \gamma \ell_1 \overline \ell_2)$. 
The black circles represent the four fermion LFV vertex, the black boxes represent the 
dipole LFV vertex, and the grey boxes represent the quarkonium bound state.}
\label{vector3body}
\end{figure}
In principle, a calculation of the amplitude ${\cal A} (V \to \gamma \ell_1 \overline \ell_2)$ involves evaluation of the 
eight diagrams shown in Fig.~\ref{vector3body}. Since the initial state is a $1^{--}$ vector meson, the contributions of the 
axial, scalar, and pseudoscalar are contained in diagrams \ref{diagram a} and \ref{diagram b}. The diagrams \ref{diagram c}
and \ref{diagram d} contain the vector and tensor operator contributions and \ref{diagram e}-\ref{diagram h} are generated 
by the dipole operator contributions. By the same arguments as above, we shall also ignore those in this paper. 

\begin{figure}
\subfigure[]{\includegraphics[scale=.63]{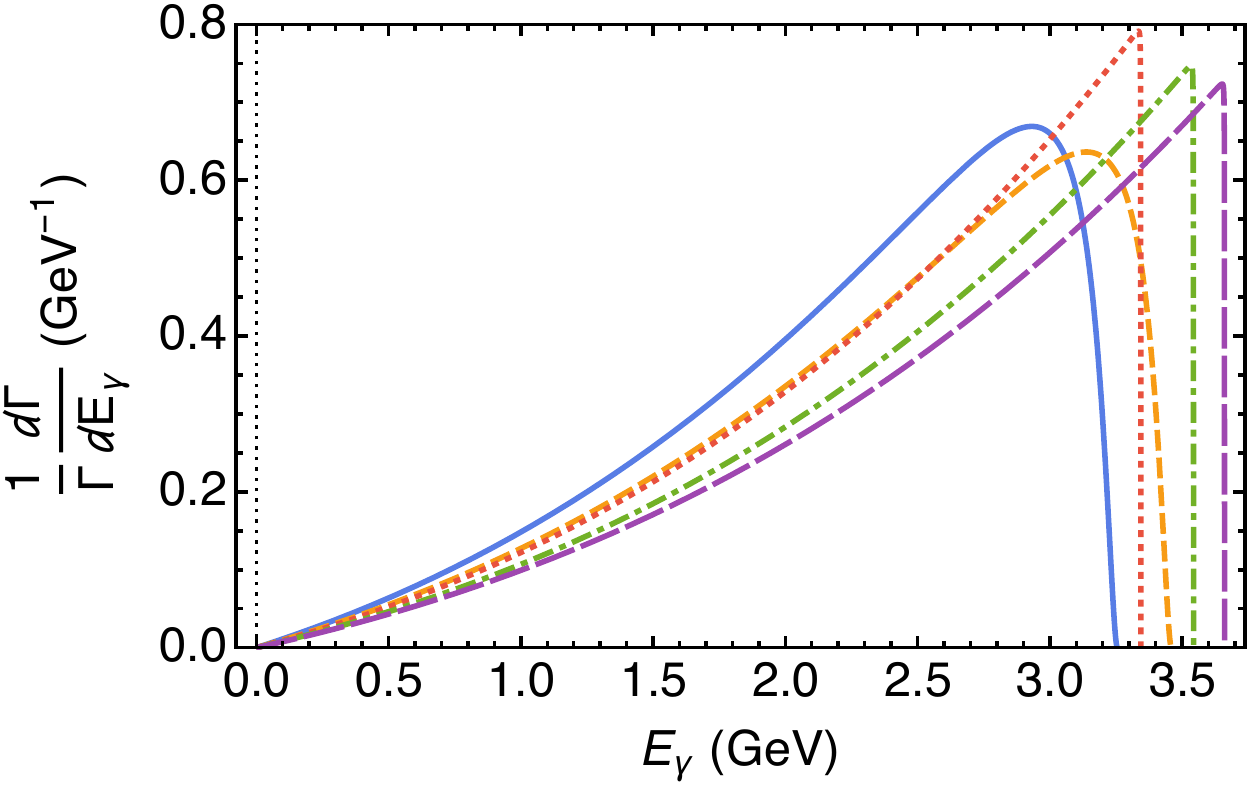} \label{diffdecayaxialbb}}
\subfigure[]{\includegraphics[scale=.63]{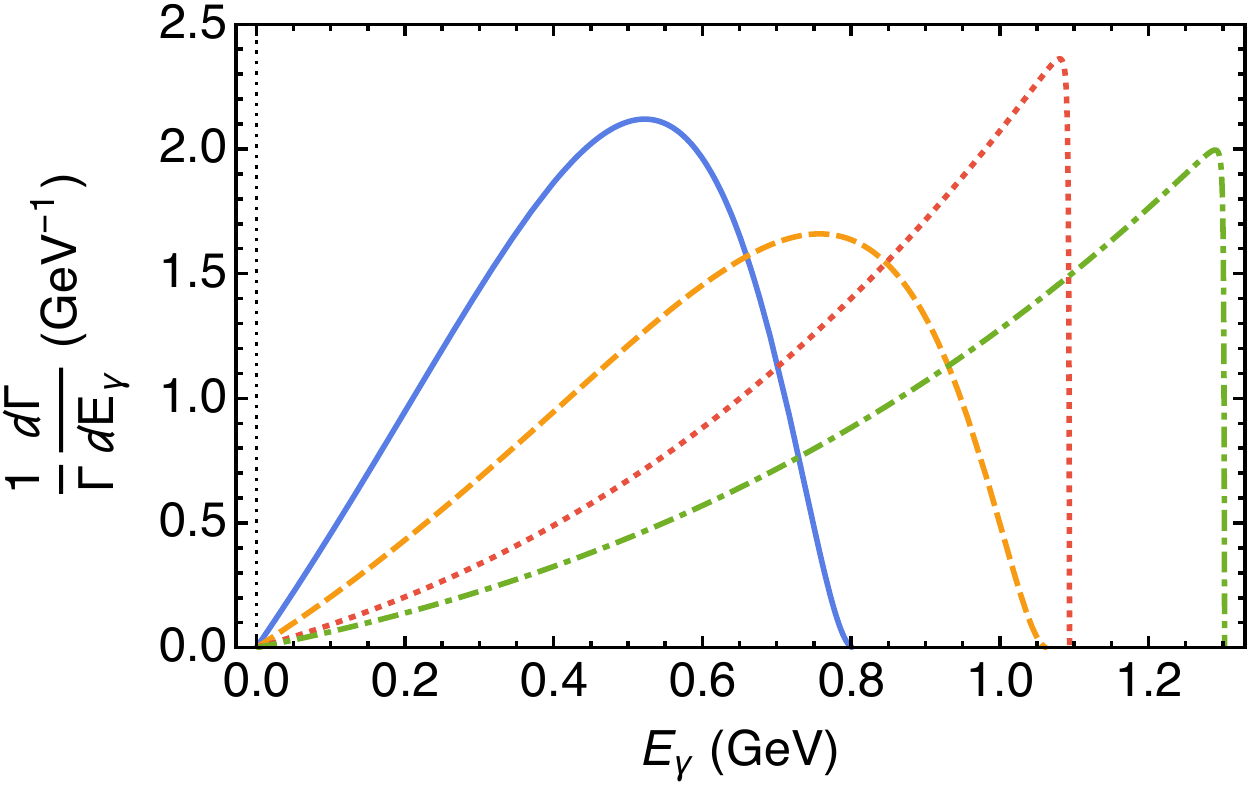} \label{diffdecayaxialcc}}
\subfigure[]{\includegraphics[scale=.63]{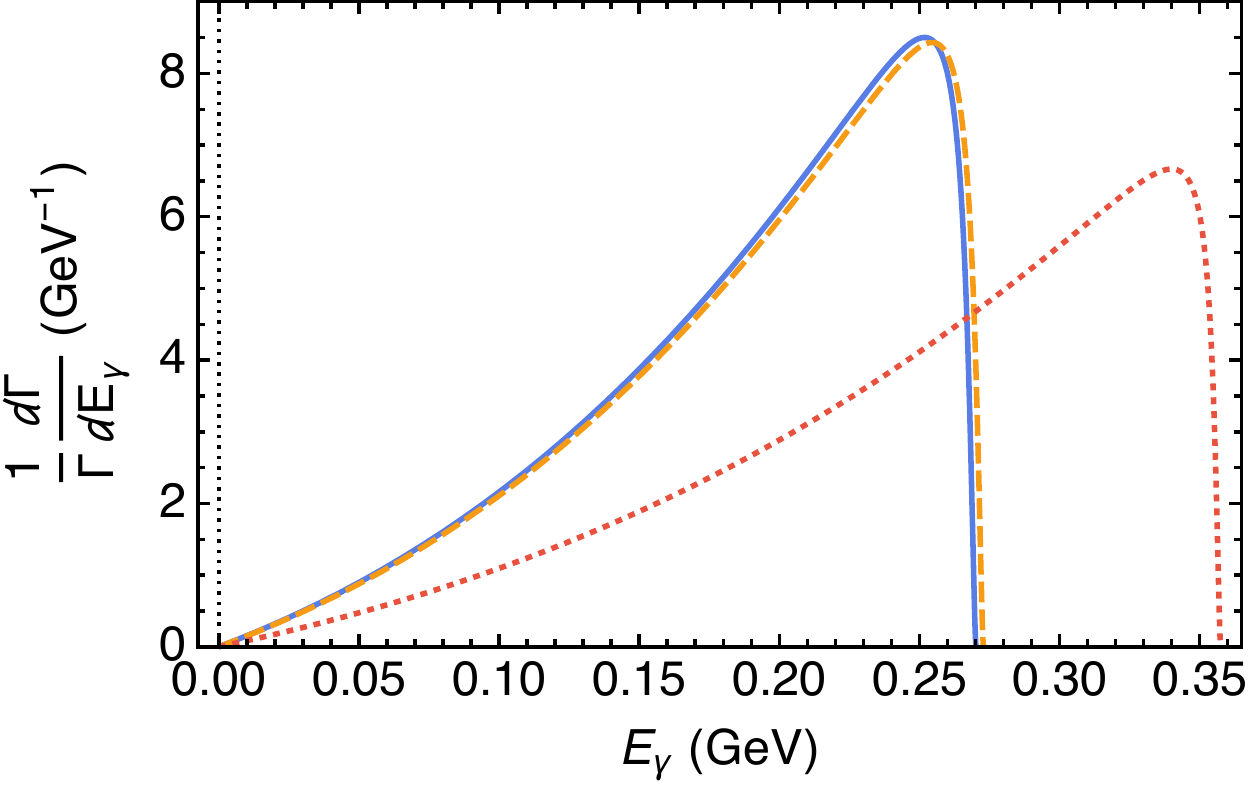} \label{diffdecayaxialqq}}
\captionsetup{singlelinecheck=off}
    \caption[]{Differential decay rates as functions of photon 
    energy $E_\gamma$ for axial operators. Plotted decay rates are for
    (a) $\Upsilon(1S) \to\gamma \mu \tau$ or $ \gamma e \tau$ (solid blue), 
    $\Upsilon(2S) \to \gamma \mu \tau$ or $\gamma e \tau$ (short-dashed gold), 
    $\Upsilon(3S) \to \gamma \mu \tau$ or $\gamma e \tau$ (dotted red),
    $\Upsilon(1S) \to \gamma e \mu$ (dot-dashed green),
    $\Upsilon(2S) \to \gamma e \mu$ and $\Upsilon(3S) \to \gamma e \mu$ (long-dashed purple);
    (b) $J\psi \to \gamma \mu \tau$ or $\gamma e \tau$ (solid blue),
    $\psi(2S) \to \gamma \mu \tau$ or $ \gamma e \tau$ (short-dashed gold), 
    $J\psi \to \gamma e \mu$ (dotted red),
    $\psi(2S) \to \gamma e \mu$ (dot-dashed green);
    (c) $\rho \to \gamma e \mu$ (solid blue),
    $\omega \to \gamma e \mu$ (short-dashed gold),
    $\phi \to \gamma e \mu$ (dotted red).}
\label{differentialdecayaxial}
\end{figure}

\begin{figure}
\subfigure[]{\includegraphics[scale=.63]{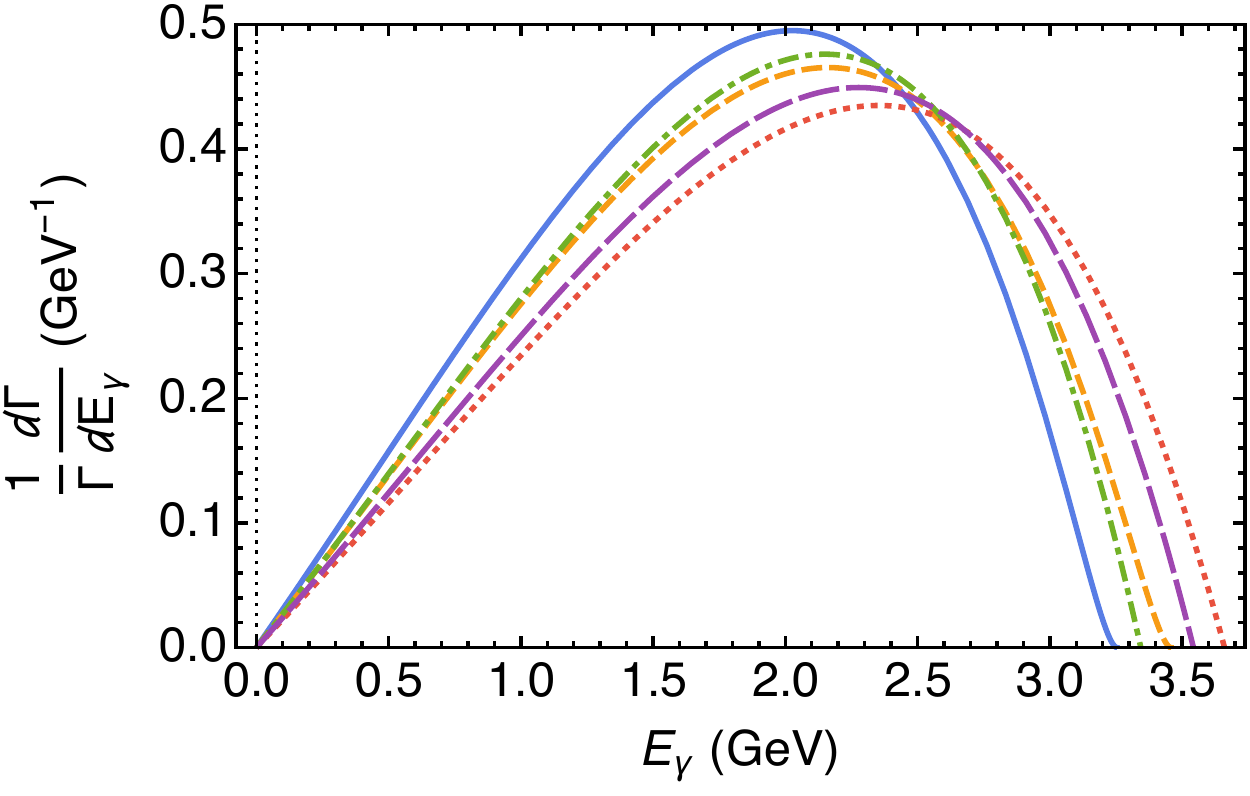} \label{diffdecayscalarbb}}
\subfigure[]{\includegraphics[scale=.63]{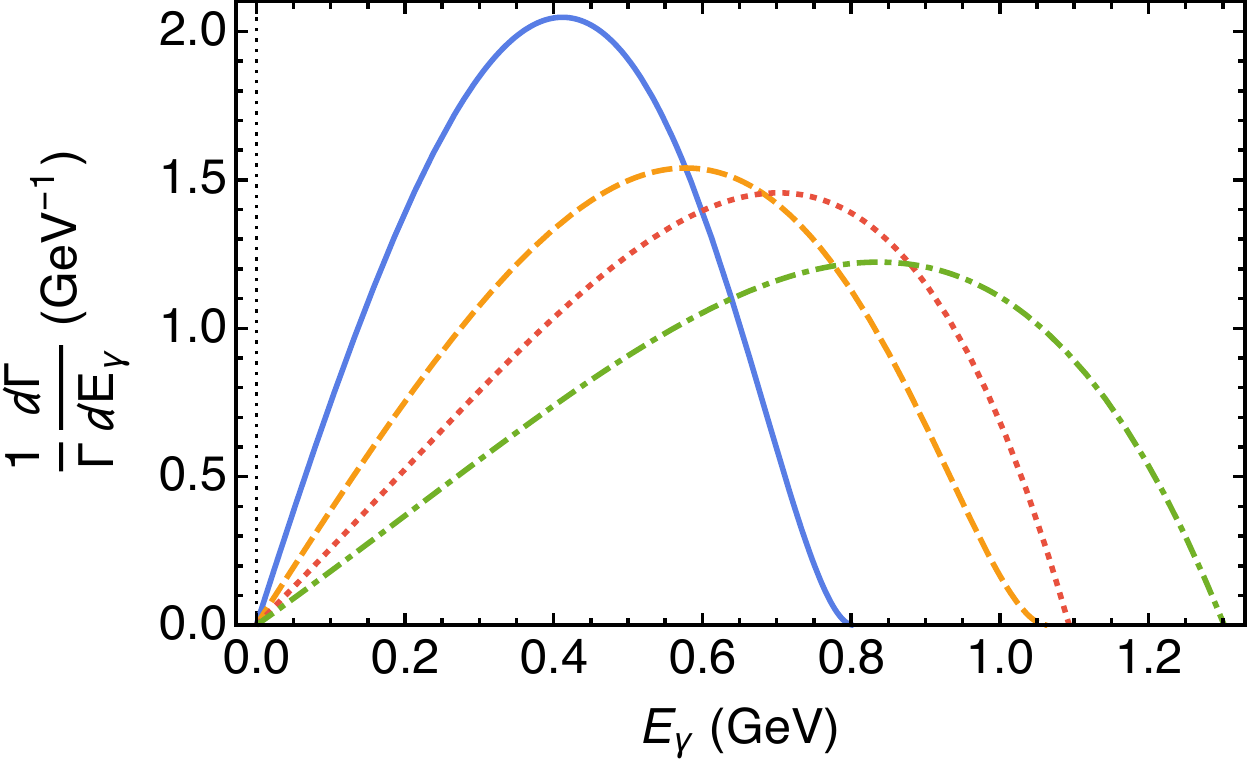} \label{diffdecayscalarcc}}
\subfigure[]{\includegraphics[scale=.63]{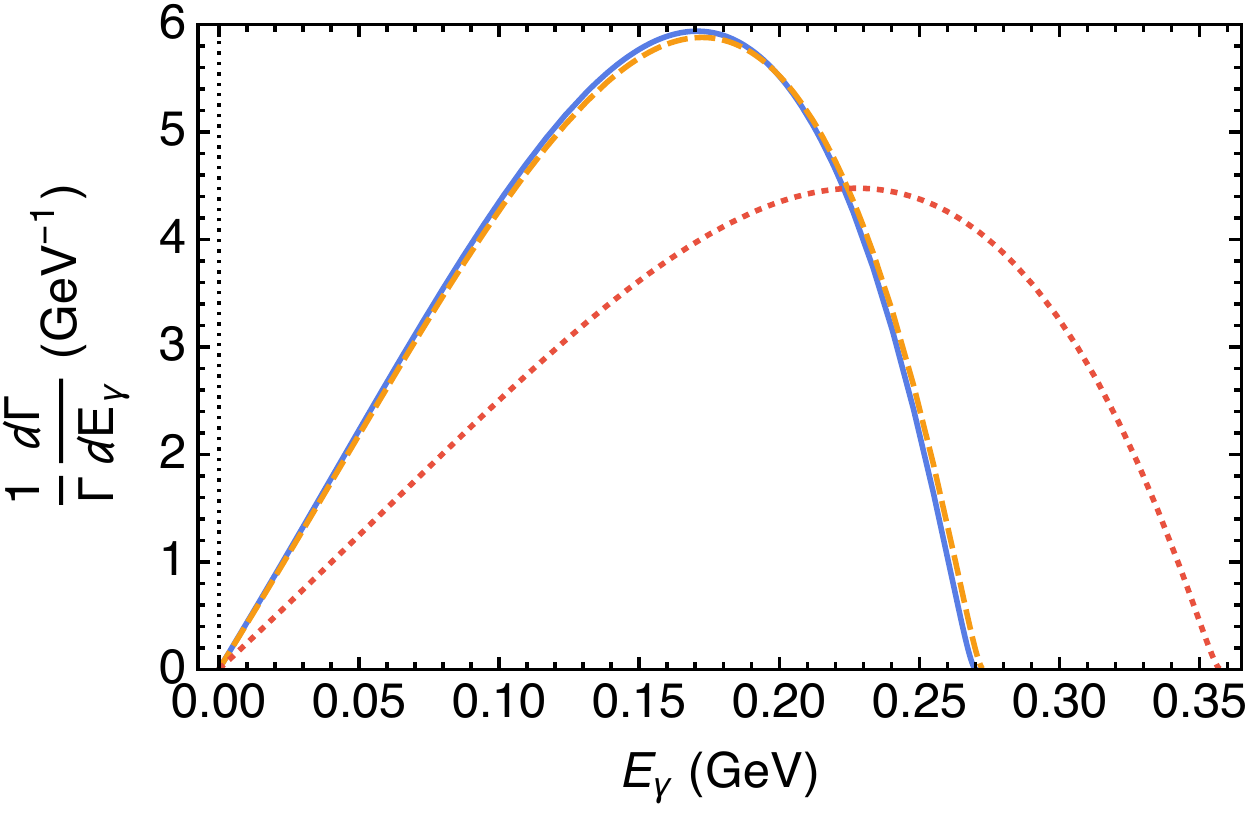} \label{diffdecayscalarqq}}
\captionsetup{singlelinecheck=off}
    \caption[]{Differential decay rates as functions of photon 
    energy $E_\gamma$ for scalar/pseudoscalar operators. Plotted decay rates are for
    (a) $\Upsilon(1S) \to \gamma \mu \tau$ or $\gamma e \tau$ (solid blue), 
    $\Upsilon(2S) \to \gamma \mu \tau$ or $ \gamma e \tau$ (short-dashed gold), 
    $\Upsilon(3S) \to \gamma \mu \tau$, $\gamma e \tau$, or $\gamma e \mu$ (dotted red),
    $\Upsilon(1S) \to \gamma e \mu$ (dot-dashed green),
    $\Upsilon(2S) \to \gamma e \mu$ (long-dashed purple);
    (b) $J\psi \to \gamma \mu \tau$ or $\gamma e \tau$ (solid blue),
    $\psi(2S) \to \gamma \mu \tau$ or $ \gamma e \tau$ (short-dashed gold), 
    $J\psi \to \gamma e \mu$ (dotted red),
    $\psi(2S) \to \gamma e \mu$ (dot-dashed green);
    (c) $\rho \to \gamma e \mu$ (solid blue),
    $\omega \to \gamma e \mu$ (short-dashed gold),
    $\phi \to \gamma e \mu$ (dotted red).}
\label{differentialdecayscalar}
\end{figure}

A calculation of ${\cal A} (V \to \gamma \ell_1 \overline \ell_2)$ presented in this paper involves a model to describe the 
quark-antiquark wave function of the quarkonium state \cite{Aditya:2012ay}. We choose to follow 
\cite{Aditya:2012ay, Dziembowski:1986dr, Szczepaniak:1990dt, Lepage:1980fj} and write it as
\begin{eqnarray}
\Psi_V = \frac{I_c}{\sqrt{6}} \Phi_V(x) \left(m_V \gamma^{\alpha}+ i p^{\beta} \sigma^{\alpha \beta} \right) \epsilon^{\alpha}(p).
\label{wavefunction}
\end{eqnarray} 
Here $I_c$ is the identity matrix in color space, $x$ is the quarkonium momentum fraction carried by one of the constituent 
quarks, and $p$ is the momentum of the vector meson. The distribution amplitude, $\Phi_V(x)$, in Eq.~(\ref{wavefunction}) 
is defined as
\begin{eqnarray}\label{distributionfunction}
\Phi_V(x) = \frac{f_V}{2\sqrt{6}}  \delta(x-1/2),  
\end{eqnarray}
where $f_V$ is a decay constant defined in Eq.~(\ref{DeConV}). We chose the simplest wave function which makes the 
approximation that each constituent quark carries half the meson's momentum, which is a good approximation for the 
heavy quark states such as $\Upsilon(nS)$ or $J/\psi$. The non-local matrix element that is relevant for the 
radiative transition is then expressed in terms of an integral over momentum fraction:
\begin{eqnarray}
\braket{0|\overline q \Gamma^{\mu} q}V =\int_{0}^{1} \text{Tr}[\Gamma^{\mu} \Psi_V] dx  \label{matrixelement}.
\end{eqnarray}
We can now calculate the total and differential decay rates. Assuming single operator dominance, the axial, scalar, and 
pseudoscalar operators lead to the following differential decay rates:
\begin{eqnarray} \label{3bodydiffdecayrates}
\frac{d\Gamma_{V \to \gamma \ell_1 \overline \ell_2}^A}{dm_{12}^2} &=& \frac{1}{9} \frac{\alpha Q_q^2}{\left(4 \pi\right)^2} 
\frac{f_V^2}{\Lambda^4} \left(C_{AL}^2+C_{AR}^2 \right) \frac{\left(m_V^2-m_{12}^2\right) 
\left(2 m_V^2 y^2 + m_{12}^2\right) \left(m_V^2 y^2 - m_{12}^2\right)^2}{m_V m_{12}^6}, 
\nonumber \\
\frac{d\Gamma_{V \to \gamma \ell_1 \overline \ell_2}^S}{dm_{12}^2} &=& \frac{1}{24} 
\frac{\alpha Q_q^2}{\left(4 \pi\right)^2} \frac{f_V^2 G_F^2 m_V}{\Lambda^4} 
\left(C_{SL}^2+C_{SR}^2 \right) \frac{y^2 \left(m_V^2-m_{12}^2\right) 
\left(m_V^2 y^2 - m_{12}^2\right)^2}{m_{12}^2}, 
\\
\frac{d\Gamma_{V \to \gamma \ell_1 \overline \ell_2}^P}{dm_{12}^2} &=& \frac{1}{24} 
\frac{\alpha Q_q^2}{\left(4 \pi\right)^2} \frac{f_V^2 G_F^2 m_V}{\Lambda^4} 
\left(C_{PL}^2+C_{PR}^2 \right) \frac{y^2 \left(m_V^2-m_{12}^2\right) 
\left(m_V^2 y^2 - m_{12}^2\right)^2}{m_{12}^2}.
\nonumber
\end{eqnarray}
Here $y$ is defined to be the same as in Sect.~\ref{Spin1LL} and we follow the usual definition of 
the Mandelstam variable $m_{12}^2 = \left(p_1+p_2\right)^2$ \cite{PDG}, where momentum $p_1$ and $p_2$ 
correspond to $\ell_1$ and $\ell_2$. Note that in writing Eqs.~(\ref{3bodydiffdecayrates}) and (\ref{3bodydecayrate}) 
we suppressed some of the indices of the Wilson coefficients (i.e. $C_{SL}^{q \ell_1 \ell_2} \to C_{SL}$) 
for brevity. The total decay rates for the RLFV transitions can be found by integrating 
Eq.~(\ref{3bodydiffdecayrates}) over $m_{12}^2$, which gives
\begin{eqnarray} \label{3bodydecayrate}
\Gamma_A (V \to \gamma \ell_1 \overline \ell_2) &=& \frac{1}{18} \frac{\alpha Q_q^2}{\left(4 \pi\right)^2} 
\frac{f_V^2 m_V^3}{\Lambda^4} \left(C_{AL}^2+C_{AR}^2 \right) f(y^2), 
\nonumber \\
\Gamma_S(V \to \gamma \ell_1 \overline \ell_2) &=& \frac{1}{144} \frac{\alpha Q_q^2}{\left(4 \pi\right)^2} 
\frac{f_V^2 G_F^2 m_V^7}{\Lambda^4} \left(C_{SL}^2+C_{SR}^2 \right) y^2 f(y^2), 
\\
\Gamma_P (V \to \gamma \ell_1 \overline \ell_2) &=& \frac{1}{144} \frac{\alpha Q_q^2}{\left(4 \pi\right)^2} 
\frac{f_V^2 G_F^2 m_V^7}{\Lambda^4} \left(C_{PL}^2+C_{PR}^2 \right) y^2 f(y^2),
\nonumber
\end{eqnarray}
where $f(y^2) = 1-6y^2-12y^4\text{log}\left(y\right)+3y^4+2y^6$.
We can use Eq.~(\ref{3bodydecayrate}) to normalize differential decay distributions, so that 
they are independent of the unknown Wilson coefficients and plot the normalized decay distributions
under the assumption of a single operator dominance. 
We show differential photon spectra in $V \to \gamma \ell_1 \overline \ell_2$ decay
in Fig.~\ref{differentialdecayaxial} for the axial operators, and in Fig. \ref{differentialdecayscalar} for 
the scalar or pseudoscalar ones.

Since no experimental constraints are available for the RLFV decays of
vector quarkonia, we cannot yet place any constraints on the Wilson coefficients from those 
transitions. 

\section{Conclusions}\label{Conclusions}

Lepton flavor violating transitions provide a powerful engine for new physics searches. 
Any new physics model that incorporates flavor and involves flavor-violating 
interactions at high energy scales can be cast in terms of the effective Lagrangian 
of Eq.~({\ref{Leff}}) at low energies. We argued that Wilson coefficients of this 
Lagrangian could be effectively probed by studying decays of quarkonium states with 
different spin-parity quantum numbers, providing complementary constraints to those
obtained from tau and mu decays \cite{Raidal:2008jk,Bruser:2015yka}.

The proposed framework allows us to select two-body quarkonium decays 
in such a way that only operators with particular quantum numbers are probed,
significantly reducing the reliance on the single operator dominance assumption that is
prevalent in constraining the parameters of the effective LFV Lagrangian.
We also argued that studies of RLFV decays could provide 
important complementary access to those effective operators.

With new data coming form the LHC experiments and Belle II experiment, 
we strongly encourage our colleagues to 
provide experimental constraints on both the LFV and RLFV transitions 
discussed in this paper.

\begin{acknowledgments}
We would like to thank Alexander Khodjamirian for  useful discussions.
This work has been supported in part by the U.S. Department of Energy 
under contract DE-SC0007983, and  by Fermilab's Intensity Frontier Fellowship.
A.A.P. is a Comenius Guest Professor at the University of Siegen. 
\end{acknowledgments}



\begin{thebibliography}{99}

\bibitem{Raidal:2008jk} 
  M.~Raidal {\it et al.},
  Eur.\ Phys.\ J.\ C {\bf 57}, 13 (2008)
  doi:10.1140/epjc/s10052-008-0715-2
  [arXiv:0801.1826 [hep-ph]].

\bibitem{Celis:2014asa} 
  A.~Celis, V.~Cirigliano and E.~Passemar,
  Phys.\ Rev.\ D {\bf 89}, no. 9, 095014 (2014)
  doi:10.1103/PhysRevD.89.095014
  [arXiv:1403.5781 [hep-ph]].

\bibitem{Petrov:2013vka} 
  A.~A.~Petrov and D.~V.~Zhuridov,
  Phys.\ Rev.\ D {\bf 89}, no. 3, 033005 (2014)
  doi:10.1103/PhysRevD.89.033005
  [arXiv:1308.6561 [hep-ph]].

\bibitem{PDG} 
  K.~A.~Olive {\it et al.}  [Particle Data Group Collaboration],
  Chin.\ Phys.\ C {\bf 38}, 090001 (2014).

\bibitem{Colquhoun:2014ica} 
  B.~Colquhoun, R.~J.~Dowdall, C.~T.~H.~Davies, K.~Hornbostel and G.~P.~Lepage,
  Phys.\ Rev.\ D {\bf 91}, no. 7, 074514 (2015)
  doi:10.1103/PhysRevD.91.074514
  [arXiv:1408.5768 [hep-lat]].

\bibitem{Abada:2015zea} 
  A.~Abada, D.~Becirevic, M.~Lucente and O.~Sumensari,
  Phys.\ Rev.\ D {\bf 91}, no. 11, 113013 (2015)
  doi:10.1103/PhysRevD.91.113013
  [arXiv:1503.04159 [hep-ph]].

\bibitem{Becirevic:2013bsa} 
  D.~Becirevic, G.~Duplancia, B.~Klajn, B.~Meli‡ and F.~Sanfilippo,
  Nucl.\ Phys.\ B {\bf 883}, 306 (2014)
  doi:10.1016/j.nuclphysb.2014.03.024
  [arXiv:1312.2858 [hep-ph]].

\bibitem{MaiordeSousa:2012vv} 
  M.~S.~Maior de Sousa and R.~Rodrigues da Silva,
  arXiv:1205.6793 [hep-ph].

\bibitem{Donald:2013pea} 
  G.~C.~Donald {\it et al.} [HPQCD Collaboration],
  Phys.\ Rev.\ D {\bf 90}, no. 7, 074506 (2014)
  doi:10.1103/PhysRevD.90.074506
  [arXiv:1311.6669 [hep-lat]].

\bibitem{Chen:2015tpa} 
  Y.~Chen, A.~Alexandru, T.~Draper, K.~F.~Liu, Z.~Liu and Y.~B.~Yang,
  arXiv:1507.02541 [hep-ph];
   V.~V.~Braguta,
  Phys.\ Rev.\ D {\bf 75}, 094016 (2007)
  doi:10.1103/PhysRevD.75.094016
  [hep-ph/0701234 [HEP-PH]].

\bibitem{Khodjamirian:2015dda} 
  A.~Khodjamirian, T.~Mannel and A.~A.~Petrov,
  JHEP {\bf 1511}, 142 (2015)
  doi:10.1007/JHEP11(2015)142
  [arXiv:1509.07123 [hep-ph]].
  
\bibitem{Petrov:2013nia} 
  A.~A.~Petrov and W.~Shepherd,
  Phys.\ Lett.\ B {\bf 730}, 178 (2014)
  doi:10.1016/j.physletb.2014.01.051
  [arXiv:1311.1511 [hep-ph]].

\bibitem{Brambilla:2010cs} 
  N.~Brambilla {\it et al.},
  Eur.\ Phys.\ J.\ C {\bf 71}, 1534 (2011)
  doi:10.1140/epjc/s10052-010-1534-9
  [arXiv:1010.5827 [hep-ph]].

\bibitem{McNeile:2012qf} 
  C.~McNeile, C.~T.~H.~Davies, E.~Follana, K.~Hornbostel and G.~P.~Lepage,
  Phys.\ Rev.\ D {\bf 86}, 074503 (2012)
  doi:10.1103/PhysRevD.86.074503
  [arXiv:1207.0994 [hep-lat]].

\bibitem{Beneke:2002jn} 
  M.~Beneke and M.~Neubert,
  Nucl.\ Phys.\ B {\bf 651}, 225 (2003)
  doi:10.1016/S0550-3213(02)01091-X
  [hep-ph/0210085].

\bibitem{Godfrey:2015vda} 
  S.~Godfrey and H.~E.~Logan,
  Phys.\ Rev.\ D {\bf 93}, no. 5, 055014 (2016)
  doi:10.1103/PhysRevD.93.055014
  [arXiv:1510.04659 [hep-ph]].

\bibitem{Godfrey:2015dia} 
  S.~Godfrey and K.~Moats,
  Phys.\ Rev.\ D {\bf 92}, no. 5, 054034 (2015)
  doi:10.1103/PhysRevD.92.054034
  [arXiv:1507.00024 [hep-ph]].

\bibitem{Aditya:2012ay} 
  Y.~G.~Aditya, K.~J.~Healey and A.~A.~Petrov,
  Phys.\ Lett.\ B {\bf 710}, 118 (2012)
  doi:10.1016/j.physletb.2012.02.042
  [arXiv:1201.1007 [hep-ph]].

\bibitem{HazardPetrovFuture} 
  D.~E.~Hazard and A.~A.~Petrov, to be published

\bibitem{Dziembowski:1986dr} 
  Z.~Dziembowski and L.~Mankiewicz,
  Phys.\ Rev.\ Lett.\  {\bf 58}, 2175 (1987).
  doi:10.1103/PhysRevLett.58.2175
  
\bibitem{Szczepaniak:1990dt} 
  A.~Szczepaniak, E.~M.~Henley and S.~J.~Brodsky,
  Phys.\ Lett.\ B {\bf 243}, 287 (1990).
  doi:10.1016/0370-2693(90)90853-X
  
\bibitem{Lepage:1980fj} 
  G.~P.~Lepage and S.~J.~Brodsky,
  Phys.\ Rev.\ D {\bf 22}, 2157 (1980).
  doi:10.1103/PhysRevD.22.2157
 
\bibitem{Bruser:2015yka} 
  R.~Bruser, T.~Feldmann, B.~O.~Lange, T.~Mannel and S.~Turczyk,
  JHEP {\bf 1510}, 082 (2015)
  doi:10.1007/JHEP10(2015)082
  [arXiv:1506.07786 [hep-ph]].
 
\end{thebibliography}
\end{document}